\documentclass[11pt]{article}
\usepackage{fullpage}
\usepackage{graphicx}
\usepackage{latexsym,amsmath,amssymb,amsthm,epic,eepic,multirow}
\usepackage{natbib}

\usepackage{multirow}
\usepackage{subfigure} % for two figures side-by-side
\usepackage{natbib}
\usepackage{algorithm}
\usepackage{algorithmic}
\usepackage{mdframed}
\usepackage{tikz}
\usepackage{hyperref}
\usepackage{listings} % for highlighting the R code
\usepackage{color} % for highlighting the R code

\definecolor{upmaroon}{rgb}{0.48, 0.07, 0.07}
\definecolor{royalazure}{rgb}{0.0, 0.22, 0.66}
\definecolor{pakistangreen}{rgb}{0.0, 0.4, 0.0}

\newcommand{\PP}{\mathbb{P}}
\newcommand{\EE}{\mathbb{E}}

\newcommand{\eps}{\varepsilon}

\newcommand{\Cov}{\mbox{Cov}}

\newcommand{\Var}{\mbox{Var}}

\newcommand{\bx}{\mathbf{X}}
\newcommand{\argmin}{\mathrm{argmin}}
\newcommand{\pa}{\mathrm{pa}}

\newtheorem{prop}{Proposition}

\usepackage{color}

\begin{document}

\title{Hierarchical inference for genome-wide association studies:\\
a view on methodology with software}  
  
\author{Claude Renaux, Laura Buzdugan, Markus Kalisch and Peter
  B\"uhlmann\\Seminar for Statistics, ETH Z\"urich}
%\date{}

\maketitle

\begin{abstract}
We provide a view on high-dimensional statistical inference for
genome-wide association studies (GWAS). It is in part a review but covers
also new developments for meta analysis with multiple studies and novel
software in terms of an \textsf{R}-package \texttt{hierinf}. Inference and
assessment of significance is based on very high-dimensional multivariate
(generalized) linear models: in contrast to often used marginal approaches,
this provides a step towards more causal-oriented inference. 
\end{abstract}

\section{Introduction}

We provide a selective review (or a view) on high-dimensional
statistical inference for genome-wide association studies (GWAS). In doing so,
we give an illustration of our own software in terms of the
\textsf{R}-package \texttt{hierinf} and we also include
some novel methodological 
aspects and results. Both of the mentioned topics, 
high-dimensional inference and GWAS, have been
rapidly evolving over the last years and we do not aim here to present a
broad overview. Instead, we focus on the combination of the two and
consider inference in a multivariate model which quantifies effects after
adjusting for all remaining single 
nucleotide polymorphism (SNP) covariates. Assigning uncertainties in such a
multiple regression model has received fairly little attention so
far, perhaps because of the difficulty to deal in practice with the very
high-dimensionality in GWAS with $p \approx O(10^6)$ SNP covariates.   

Univariate approaches for significance of a SNP being marginally associated to a
response variable (sometimes denoted as phenotype) have been widely adopted in
the last decades. The main challenge with such marginal approaches is the
multiple testing adjustment:  
the false discovery rate (FDR) \citep{benjamini95}
has become very popular as an error criterion which is
less conservative than the familywise error rate (FWER), see for example
\citet{storey03, sabatti03, benjamini05}.
\citet{peterson16} consider a hierarchical formulation for the FDR: their
hierarchical 
procedure is very different though than the hierarchical inference scheme
which we propagate in this article (in their work, the hierarchy originates
from having multiple phenotypes; in contrast, the hierarchy in our
approach addresses the issue of highly correlated covariates in a
(generalized) linear model). 

There are several proposals which consider a multivariate regression
model. 
\citet{Baierl06} consider model selection in low-dimensional QTL modeling
which is not of high-dimensional nature. 
\citet{frommlet12} and \citet{dolejsi14} have further developed 
the methodology from \citet{Baierl06} and applied it to real GWAS data, 
i.e. in the high-dimensional context.
Furthermore for GWAS, Bayesian approaches
\citep{hoggart08,carbonetto12}, Ridge regression 
\citep{maloetal08} or the   
Lasso for screening important covariates (including interactions) 
\citep{wuetal10,wuroed10} have been considered and used, further proposals
include a combination of the Lasso and linear mixed models
\citep{raketal13,zhouetal13}, the Bayesian Lasso 
\citep{lietal11} or stability selection for sparse estimators
\citep{alexlange11,helin11}. None of these proposals compute frequentist
p-values for single or groups of SNPs, but methods based on stability
selection lead to control of the number of false positives
\citep{mebu10}. More recently, interesting work has been pursued for control
of the FDR after selection in a multivariate regression model
\citep{brzyski17}.  
The proposed
procedure is first pre-screening for the level of resolution to identify
regions or groups of SNPs and then controls the FDR for the pre-screened
first-stage regions, see also \citet{helleretal2017} when using marginal
tests. In a sense, the work by \citet{brzyski17} comes closest to our proposal 
with a hierarchical structure: both approaches share the point that
for GWAS, data-driven aggregation of hypotheses results in more power. For
an overview of univariate and multivariate methods which have been
published until the year of 2015,
we also refer to the monograph by \citet{frommletetal16}.  

We have recently proposed high-dimensional hierarchical inference for
assigning statistical 
significance in terms of p-values for groups of SNPs being associated to a
response variable: \cite{buzduganetal16}
considers this approach for human GWAS and \cite{klasenetal16} for GWAS
with plants. The methodological and theoretical concepts have been worked
out in \cite{manpb16} and \cite{manpb16b}. The hierarchy enables in a
fully data-driven way to infer significant groups or
regions of SNPs at an adaptive resolution, by controlling the familywise
error rate (FWER). Although the FWER seems overly conservative, we still
detect small groups of SNPs in real datasets. We will review this approach
and extend it to the setting of multiple studies using concepts of
meta-analysis. The difference to pre-screening and selection techniques
as in e.g. \citet{brzyski17}
is that we do not need to choose the amount 
of pre-screening at the beginning: the entire procedure is fully
data-driven, leading to high resolution (small groups of SNPs) if the signal
is strong in relation to the strength of correlation among the SNPs and
vice-versa yielding low resolution if the signal is weak. The hierarchy
itself is constructed by either  
clustering the SNPs according to their strength of squared correlations or
by partitioning of the genomic sequence into blocks of consecutive genomic
positions corresponding to groupings of the SNPs. Our
procedure is based on an efficient hierarchical multiple testing adjustment
from \citet{Meins08}. The power of a sequential approach for controlling
the FWER has been thoroughly discussed in \citet{GoSo2010}. 
The scheme by \citet{meijer15} could be an interesting
alternative when looking at region-based groups of SNPs: it is more
flexible at the price of a higher multiple testing adjustment and thus, it
is unclear whether it would exhibit more power. 

\section{High-dimensional hierarchical inference}\label{sec.highdiminfer} 

We build the statistical inference on a multiple regression model where all
the measured SNPs enter as covariates in the model. We will mainly focus
on a linear model:
\begin{eqnarray}\label{lin.mod}
Y = \mu + \bx \beta^0 + \eps,
\end{eqnarray}
with $n \times 1$ response vector $Y$, $n \times p$ design matrix $\bx$ and
$n \times 1$ vector of stochastic errors $\eps$ and intercept $\mu$. The
superscript ``$^0$'' denotes the ``true'' underlying parameter of the
data-generating distribution. We usually
assume fixed design and i.i.d. errors with $\EE[\eps_i] = 0,\ \Var(\eps_i)
= \sigma^2$. We denote the $i$th row and the $j$th column of $\bx$ by $X_i$
and $X^{(j)}$, respectively. The assumption about fixed design is not
really a loss of generality as long as the linear model is correct: if the
covariates are random, we can always condition on $\bx$ (and the linear
model is still correct) and perform the statistical inference conditional
on $\bx$.  

\paragraph{Genome-wide association study.}
In a GWAS, the covariates corresponding to the columns of $\bx$ are the
SNPs.  The response variable can be continuous like e.g. a growth rate of a
plant or binary encoding the status ``healthy'' or
``diseased'' (see Section \ref{subsec.hierinfrealdatasets} for some
examples). For the latter, we would then consider a
logistic regression model as in \eqref{glm.mod}. 
In general, the sample size is about
$n \approx 
3'000$ whereas the number of SNP covariates is in the order of $p \approx
10^6$. Obviously the model
in \eqref{lin.mod} in the setting of GWAS is very high-dimensional with
many more unknown parameters than sample size, i.e., $p \gg n$. 

We note that the multiple model in \eqref{lin.mod} is very different from a
marginal model  
\begin{eqnarray*}
Y = \mu_j + \gamma_j X^{(j)} + \tilde{\eps}^{(j)},
\end{eqnarray*}
where the response variable is modeled for every covariate $X^{(j)}$
individually. The marginal model does not take into account how much of an
effect is due to other covariates (i.e., the marginal effect is not
adjusted for other covariates), and the multiple regression model is much
more powerful towards causal inference as discussed in Section
\ref{subsec.causal}.  

\bigskip
The model has to be adapted if the response is not continuous. 
The response could be a binary variable encoding a disease status, e.g. if a 
patient has diabetes or is healthy.
We can extend the methodology to generalized linear
models of the form
\begin{eqnarray}\label{glm.mod}
& &Y_i\ \mbox{independent with},\nonumber\\
& &\eta_i = g(\EE[Y_i]) = \mu + \sum_{j=1}^p \beta^0_j X_i^{(j)},
\end{eqnarray}
where $g(\cdot)$ is a real-valued link function. The most prominent example
which we will use is a logistic regression model, where $Y_i \in \{0,1\}$
is binary, $\pi_i = \pi_i(X_i) = \PP[Y_i=1|X_i]$ and link function $g(\pi)
= \log(\pi)/\log(1-\pi)$.
We will illustrate such an extension for GWAS
analysis in Section \ref{subsec.hierinfrealdatasets}. 

In the sequel, for simplicity, we usually consider a linear
model. The extension of the methodology and computations for generalized
linear models is straightforward and the case of a logistic model is
implemented in our software \texttt{hierinf}, which is an \textsf{R} 
package available on bioconductor, as described in Section
\ref{sec.hierinf}. The theoretical results which we review in the 
subsequent sections carry over to generalized linear 
models: the underlying analysis is more delicate though, see for example
\citet{pbvdg11}. 

\subsection{High-dimensional inference} 

A first goal is to infer the very many unknown regression 
parameters $\beta^0$ in \eqref{lin.mod} or \eqref{glm.mod}, respectively. 
This means that we are interested in estimating the regression coefficients 
in one of the afore mentioned models. A next important aim is to 
perform statistical hypothesis testing, which is described in Section 
\ref{subsec.testing}.

Because of the high-dimensionality of the problem at hand, the estimated
regression  
parameters $\hat{\beta}$ are regularized and enforced to be sparse,
i.e. many of its components are equal to zero. 
We restrict ourselves
for the moment to the case of a linear model \eqref{lin.mod}.
The Lasso \citep{tibs96} has become a very popular tool for point estimation: 
\begin{eqnarray}\label{lasso}
\hat{\beta}(\lambda) = \argmin_{\beta}(\|Y - \bx \beta\|_2^2/n + \lambda
  \|\beta\|_1), 
\end{eqnarray} 
where $\lambda > 0$ is a regularization parameter which needs to be
chosen. The Euclidean or $\mbox{L}_2$ norm is denoted by $\| \cdot
  \|_2$ and the Manhattan or $\mbox{L}_1$ norm by $\| \cdot \|_1$. 
The first term in the above equation, the sum of squared 
residuals, is identical to the case of least squares estimation for a  
low-dimensional regression problem. The sum of squared residuals is divided
by the number of observations $n$ in order to achieve a proper scaling but
it does not change the methodology. The second term penalizes the size of
the regression parameters: because of the ``geometry'' of the $\mbox{L}_1$-norm,
Lasso is a sparse estimator with many components being exactly equal to
zero (depending on the value of $\lambda$). 

The columns or covariables of the $n \times p$ design matrix
$\bx$ are denoted as before by $X^{(j)}$ with $j = 1, \ldots, p$.
Here, the $Y$ response and all the covariates $X^{(j)}$ are assumed
to be mean centered so that the intercept $\mu$ can be dropped from the
model. This is a convenient way to estimate the unknown parameters
$\beta^0$. Furthermore, the Lasso \eqref{lasso} usually makes most sense if
all the covariates are on the same scale, as implemented per default in
the \textsf{R}-package \texttt{glmnet} \citep{friedman10}. 
The penalty term penalizes all the variables with the same amount 
which only makes sense if they are standardized. 
For GWAS, the SNP covariates take values 
in $\{0,1,2\}$ (minor allele frequencies) and are treated as  
numerical values: since they are ``on
the same scale'', we do not standardize them to equal standard
deviation.
Treating them as numerical or continuous rather than categorical or
ordinal variables has the advantage of using only one parameter for
each SNP covariate, whereas a categorical approach with main effects or
full interactions would require $2\cdot p$ or $3^p-1$ parameters,
respectively. Using a continuous scale modeling for SNPs is a rather common
approach, see for example \citet{cantor10} or \citet{bush12}. This means
that we are searching for additive effects. One typically has 
reasonable power to detect additive and dominant effects whereas  
for recessive effects the study might be underpowered \citep{bush12}.

Throughout the paper, whenever we will make some asymptotic statements,
they are meant to be that the dimension $p$ as well as the sample size $n$
tend to infinity, i.e., we adopt a ``changing model'' (sometimes called
``triangular array'') asymptotics. That is, the dimension $p = p_n$
and the model parameters $\beta = \beta_n$, $\mu = \mu_n$ and $\sigma =
\sigma_n$ (for linear models) depend on $n$ and typically the ratio
$p_n/n \to \infty$ as $n \to \infty$.

\subsubsection{Statistical properties of the Lasso}\label{subsec.property}

\paragraph{An executive summary.} The statistical properties of the Lasso
in \eqref{lasso} have been extensively studied during the last decade. The
Lasso is a nearly optimal method for prediction and parameter estimation
when making the main assumptions on sparsity of the parameter vector
(assumption (A1) below) and identifiability in terms of
``well-posedness'' of the design matrix (assumption (A2) below). For
accurate selection of the active set of variables (having non-zero
regression coefficients), one necessarily needs a ``beta-min'' condition
(assumption (A3) below) which requires that the non-zero regression
coefficients are sufficiently large. In addition, one would necessarily
need a rather strong irrepresentable condition on the design matrix: this
can be avoided guaranteeing instead a variable screening property. The
latter is most useful in practice, and in fact a standard workhorse in many
applications, allowing to screen for the important variables and achieving
a drastic dimensionality reduction in terms of the original variables. 

\bigskip
\noindent
The two main assumptions leading to good or near optimal properties of the
Lasso for (point) estimation of $\beta^0$ are a sparsity assumption on
the parameter vector $\beta^0$ and an identifiability assumption
on the design $\bx$. The Lasso itself is a sparse estimator and hence it is
expected that it leads to good performance if the true underlying parameter
$\beta^0$ is sparse as well: the support of $\beta^0$, sometimes also
called the active set, is denoted by 
\begin{eqnarray*}
S_0 = \{j;\ \beta^0_j \neq 0\},
\end{eqnarray*}
and we will assume that its cardinality $s_0 = |S_0|$ is smaller than
$\mathrm{rank}(\bx) \le n$.
Regarding identifiability, since
$\mathrm{rank}(\bx) \le n < p$, the null-space of $\bx$ is 
not trivial and we can write
\begin{eqnarray*}
\bx \beta^0 = \bx \theta\ \mbox{for $\theta = \beta^0 + \xi$ with any $\xi$
  in the null-space of $\bx$}. 
\end{eqnarray*}
Thus, in order to estimate $\beta^0$ we must make an additional
identifiability assumption on the design $\bx$ which again relies on
sparsity with a not too large set $S_0$. 

The main assumptions are as follows:
\begin{description}
\item[(A1)] {\bf Sparsity:} The cardinality of the support or active set of
  $\beta^0$ satisfies
\begin{eqnarray*}
s_0 = |S_0| = o(a_n),\ a_n \to \infty,
\end{eqnarray*}
with typical values being $a_n = n/\log(p)$ or $a_n = \sqrt{n/\log(p)}$,
see for example \citet[eq.(2.22)]{pbvdg11}. 

\item[(A2)] {\bf Compatibility condition \citep{vandeGeer:07a}:} An
  identifiability assumption on the design $\bx$.  

For some $\phi_0 > 0$ and for
  all $\beta$ satisfying $\|\beta_{S_0^c}\|_1 \le 3 \|\beta_{S_0}\|_1$ it
  holds that 
\begin{eqnarray*}
\|\beta_{S_0}\|_1^2 \le (\beta^{\top} \hat{\Sigma} \beta) s_0/\phi_0^2,
\end{eqnarray*}
where $\hat{\Sigma} = \bx^{\top} \bx/n$ and $\beta_S$, for an index set $S
\subseteq 
\{1,\ldots ,p\}$, has elements set to zero outside the set $S$, i.e.,
$(\beta_S)_j = 0\ (j \notin S)$ and $(\beta_S)_j = \beta_j\ (j \in S)$. The
value $\phi_0>0$ is called the compatibility constant. 
\end{description}  

Assuming conditions (A1) and (A2) (with the compatibility constant
$\phi_0$) one can establish an oracle inequality of
the following form, see for example \citet[Th.6.1]{pbvdg11}. Consider a
linear model as in \eqref{lin.mod} with fixed design $\bx$, Gaussian or
sub-Gaussian errors $\eps$ and when 
using the Lasso \eqref{lasso} with regularization parameter $\lambda \asymp
\sqrt{\log(p)/n}$: 
\begin{eqnarray*}
\|\bx(\hat{\beta}(\lambda) - \beta^0)\|_2^2/n + \lambda
  \|\hat{\beta}(\lambda) - \beta^0\|_1 
  \le O_P(\lambda^2 s_0/\phi_0^2). 
\end{eqnarray*}
The parameter $\lambda$ cannot be chosen smaller than of the order
$\sqrt{\log(p)/n}$ since otherwise, the probability in the ``$O_P(\cdot)$''
notation would not become large and the statement would not hold
anymore. When choosing $\lambda \asymp \sqrt{\log(p)/n}$ and assuming that
the compatibility constant $\phi_0 \ge L > 0$ is bounded away from zero, we
obtain for 
\begin{eqnarray}
& &\mbox{prediction:}\ \|\bx(\hat{\beta} - \beta^0)\|_2^2/n \le
  O_P(s_0 \log(p)/n),\label{prederror}\\
& &\mbox{parameter estimation:}\ \|\hat{\beta} - \beta^0\|_1 \le O_P(s_0
    \sqrt{\log(p)/n}).\label{l1esterror}
\end{eqnarray}
Here, we have dropped the dependence of $\hat{\beta}$ on $\lambda$. 
The second statement is more relevant for inferring the true underlying
$\beta^0$. In particular, it is straightforward to derive a screening
property as discussed next. 

The Lasso, being a sparse estimator, is often used as a variable selection
and screening tool. We denote by 
\begin{eqnarray*}
\hat{S}(\lambda) = \{j;\ \hat{\beta}_j(\lambda) \neq 0\}.
\end{eqnarray*}
The aim would be that $\hat{S} \approx S_0$, which is a highly ambitious
goal (see below). Clearly, to infer the active set from data, the regression
coefficients in $S_0$ must be sufficiently large. This can be ensured by an
additional ``beta-min'' condition: 
\begin{description}
\item[(A3)] $\min\{|\beta^0_j|;\ \beta^0_j \neq 0\} \, = \, \min_{j \in S_0}
  |\beta^0_j| \, \ge \, C(s_0,p,n)$, \, where $C(s_0,p,n) \asymp \sqrt{s_0
    \log(p)/n}$.
\end{description} 
Assuming (A1), a slightly stronger version than (A2) in
terms of a restricted eigenvalue condition \citep{brt09}, and
(A3), we have the following screening result. For a linear model as in
\eqref{lin.mod} with fixed design $\bx$, Gaussian or sub-Gaussian errors
$\eps$ and when using the Lasso \eqref{lasso} with regularization
parameter $\lambda \asymp \sqrt{\log(p)/n}$: 
\begin{eqnarray}\label{screening}
\PP[\hat{S} \supseteq S_0] \to 1\ (p \ge n \to \infty).
\end{eqnarray}
When using the weaker compatibility condition (A2), we would then require
the beta-min condition with a larger $C(s_0,p,n) \asymp s_0
\sqrt{\log(p)/n}$. This is an immediate consequence of \eqref{l1esterror}. 

The variable screening property is a highly efficient dimension reduction
technique in terms of the original covariates. Because it holds that
$|\hat{S}(\lambda)| \le \min(n,p)$ for all $\lambda$ (assuming (A1) and
(the weaker version) of (A2)), and the latter equals 
$n$ in the high-dimensional regime with $p \gg n$, we can greatly
reduce the dimension without losing an active variable from $\hat{S}_0$.
Obviously, it would be even better if variable selection would
consistently estimate the true underlying active set, 
\begin{eqnarray*}
\PP[\hat{S}(\lambda) = S_0] \to 1\ (p \ge n \to \infty).
\end{eqnarray*}
However, such a consistent variable selection property necessarily requires
a much stronger so-called irrepresentable condition on the design
$\bx$ than the assumption in (A2) \citep{mebu06,zou06,zhaoyu06}. 

\paragraph{Practical considerations.} For the task of inference described
in Section \ref{subsec.multsamplspl} below, we aim for a
regularization parameter $\lambda$ such that the screening property
\eqref{screening} holds, i.e., that $\hat{S}(\lambda) \supseteq S_0$ holds in a
reliable way. Choosing the regularization parameter by cross-validation (by
default 10-fold 
CV), denoted by $\lambda_{\mathrm{CV}}$  typically leads to a good set
$\hat{S}(\lambda_{\mathrm{CV}})$ in comparison to 
other values of $\lambda$. 
It isn't true that there is a monotone
relationship between $\lambda$ and $\hat{S}(\lambda)$ and thus, a smaller
value $\lambda < \lambda_{\mathrm{CV}}$ does not
necessarily lead to a superset $\hat{S}(\lambda) \supseteq
\hat{S}(\lambda_{\mathrm{CV}})$. 
\citet{pbmand13} illustrate the success of various variable screening
methods with respect to true and false positives, without considering the
issue of choosing a good regularization parameter: overall, the Lasso leads
to a competitive performance in comparison to other methods. It is not so
unlikely though that the property $\hat{S} \supseteq S_0$ can be rather
far from being entirely correct: it is rare that all of the variables in
$S_0$ are contained in $\hat{S}(\lambda)$ but hopefully a
reasonable good sized fraction of $S_0$ is contained in the set $\hat{S}$
from the Lasso.  

\paragraph{The assumptions in the context of GWAS.} We discuss here
whether the theoretical assumptions hold at least approximately in the
context of GWAS. The assumption (A1) is about sparsity: it is a
speculation whether the true underlying biological phenomena are sparse:
the model is always a simplification and a best sparse approximation,
achieved by the Lasso, is often still very useful. More details about
best sparse approximation properties and weak sparsity are given in
\citet{pbvdg11} and \citet{vdg16}. 
Assumption (A2) can be justified as follows: assume that the
covariates are i.i.d. sampled from a population distribution with
covariance matrix $\Sigma$, having smallest eigenvalue bounded away from
zero. Then, if the population distribution is e.g. sub-Gaussian, the
condition (A2) holds with high probability for sparse sets $S_0$ \citep[
Cor.6.8]{pbvdg11}. It seems quite plausible that the population distribution in
a GWAS context has spatially decaying covariance behavior such that the
smallest eigenvalue is bounded away from zero, e.g. for a Toeplitz matrix
model. The main assumption is again sparsity for the set $S_0$ as in
(A1). Assumption (A3) is severe and not realistic to hold exactly in many
applications: however, it is not required for \eqref{prederror} and
\eqref{l1esterror}, and it can be avoided also for hypothesis testing as
pointed out in Section 
\ref{subsec.despars}: our proposed multi sample splitting procedure in
this paper has no theoretical guarantees without (a weaker form of) (A3)
but thanks to multiple sample splitting and averaging, it still performs
empirically reasonably well in absence of condition (A3), see for example
\citet{dezetal15}.

\subsection{Statistical hypothesis testing}\label{subsec.testing}

Our main goal is to provide p-values for statistical hypothesis tests. We
consider the following null and alternative hypotheses for the regression
parameters in the model \eqref{lin.mod} or \eqref{glm.mod}. For individual
    variables 
\begin{eqnarray*} 
H_{0,j}:\ \beta^0_j = 0\ \ \mbox{versus}\ \ H_{A,j}:\ \beta^0_j \neq 0,
\end{eqnarray*}
or for a group $G \subseteq \{1,\ldots ,p\}$ of variables:
\begin{eqnarray}\label{H0G}
H_{0,G}:\ \beta^0_j = 0\ \mbox{for all $j \in G$\ \ versus}\ \ H_{A,G}:\
  \mbox{there exists $j \in G$ with $\beta^0_j \neq 0$}. 
\end{eqnarray}
The challenge is to construct p-values in the very high-dimensional
setting with $p \gg n$ which control the error rate of falsely rejecting the
null-hypothesis (the type I error rate). There is also a computational
difficulty involved and the methods from Section \ref{subsec.despars} are 
not feasible in the context of GWAS with $p \approx 10^6$ covariates. 
And finally, there is the issue of multiple testing: this is addressed 
in Section \ref{sec.hierarchical} advocating a very powerful hierarchical 
approach. 

\subsubsection{Multi sample splitting and aggregation of p-values}\label{subsec.multsamplspl}

\paragraph{An executive summary.} Sample splitting and its improved version
of multiple sample splitting 
\citep{memepb09} is rather straightforward and, as a modular technique, it
is easy to implement. It is justified to yield valid p-values which control
(possibly conservatively) the type I error rate under the assumptions
(A1)--(A3): while (A1)--(A2) are essentially unavoidable, the beta-min
assumption (A3) is rather unpleasant since the p-value or statistical test
itself is a method to investigate whether a regression coefficient is
``smallish'' or sufficiently large (while (A3) is simply assuming the
latter). However, the method has been empirically found to be rather
reliable to control the type I error rate and yet having often reasonable 
power \citep{dezetal15} to detect a variety of alternative hypotheses. 
From a computational view
point, the procedure is scaling very nicely for very high-dimensional
problems making it feasible to be used for GWAS with $p \approx 10^6$.   

\bigskip
\noindent
The idea of the procedure is as follows. We do variable screening with an estimated set of
variables $\hat{S}$ such that \eqref{screening} holds, at least in an
approximate sense. We can then use standard low-dimensional inference
methods based on the selected variables from $\hat{S}$ only. To avoid to
use the data twice for screening and inference, we split the dataset into two
halves: select or screen for variables in the first half and pursue the
inference on the second remaining part of the dataset. This procedure is
implicitly given in the work by \citet{WR08}. 

\begin{center}
Sample splitting for p-values
\end{center}

\vspace*{-5mm}
\begin{enumerate}
\item Randomly split the sample into two parts of equal size. Denote the
  corresponding indices by $I_1,
  I_2$ with $I_i \subset \{1,\ldots ,n\}\ (i=1,2)$ such that $I_1 \cap I_2
  = \emptyset$, $I_1 \cup I_2 = \{1,\ldots ,n\}$ and $|I_1| =
  \lfloor n/2 \rfloor,\ |I_2| = n - \lfloor n/2 \rfloor$. 
\item Do variable selection or screening with the Lasso based on data with
  samples from
  $I_1$: denote the selected variables by $\hat{S}_{I_1}$. (The Lasso can
  be used for linear or also generalized linear models). We use the
    regularization such that $\hat{S}_{I_1}$ consists of the
    first $\lfloor n/6 \rfloor$ variables entering in the Lasso
    regularization path.
\item Derive p-values for individual or group hypotheses based on data with
  covariates from $\hat{S}_{I_1}$ and samples from $I_2$. Since 
  $|\hat{S}_{I_1}| = \lfloor n/6 \rfloor$ and assuming that
  $\mathrm{rank}(\bx_{I_2,\hat{S}_{I_1}}) = |\hat{S}_{I_1}| = \lfloor n/6
  \rfloor$ we can use 
  classical techniques based on least squares or likelihood ratio testing. 

For the linear regression model \eqref{lin.mod} we use for a
single variable $j \in \{1,\ldots ,p\}$,
\begin{eqnarray*}
& &\mbox{if $j \in \hat{S}_{I_1}$:}\ \mbox{p-value}\ P_j\ \mbox{from the
    two-sided t-test for $H_{0,j}$ 
  based on $(Y_{I_2},\bx_{I_2,\hat{S}_{I_1}})$};\\
& &\mbox{if $j \not \in \hat{S}_{I_1}$: set $P_j = 1$}.
\end{eqnarray*}
Similarly, for a group $G \subseteq \{1,\ldots ,p\}$,
\begin{eqnarray*}
& &\mbox{if $G \cap \hat{S}_{I_1} \neq \emptyset$:}\ \mbox{p-value}\ P_G\
    \mbox{from the partial F-test for $H_{0,\tilde{G}}$, where $\tilde{G} =
    G \cap \hat{S}_{I_1}$};\\
& &\mbox{if $G \cap \hat{S}_{I_1} = \emptyset$: set $P_G = 1$}.
\end{eqnarray*}
\end{enumerate}  
For a generalized linear model \eqref{glm.mod} we use the likelihood ratio
test instead of the t- or partial F-test.

\medskip

The sample splitting method is valid and controls the type I
error if the screening property $\hat{S} \supseteq S_0$ holds. This is due
to the fact that we have all the relevant variables in the model in the second
inference step based on data from $I_2$. The requirement for the screening
property can be a bit relaxed as analyzed in \citet{pbmand13}, allowing
also for not too many small non-zero regression coefficients. 

Note that if the intersection between a given group $G$ and the selected 
set of variables with Lasso $\hat{S}_{I_1}$ based on a half-sample is
empty, then the p-value is set to the value one. For some given large group,  
the intersection between this group and the selected set of variables from
Lasso has cardinality at most equal to $\hat{S}_{I_1}$ which is
  bounded by the half-sample size $\lfloor n/2 \rfloor$. In particular, not
  all the variables of such a group $G$ are considered for calculating the
  p-value in the other half-sample $I_2$. This works fine since we assume
that the screening property $\hat{S}_{I_1} \supseteq S_0$ of the Lasso  
holds, which implies that we control for all the relevant variables.

Unfortunately, sample splitting very much
depends on how the dataset is split into two parts, e.g., the random choice
of partitioning the data into two groups. 
To avoid this dependence on how the dataset is split, one can do the sample
splitting and inference procedure many times (e.g. 100 times) and then
aggregate the corresponding p-values in a way so that the type I error is
controlled. This aggregation step requires special attention and is
detailed below in \eqref{aggreg}. The method has been invented by
\citet{memepb09} and works as follows.

\paragraph{Multiple sample splitting for p-values.}
The multiple sample splitting approach uses the steps 1.-3. from the sample
splitting procedure above $B$
times. For a group of variables $G$, including the case of $G = \{j\}$
being a singleton, this leads to $B$ p-values
\begin{eqnarray*}
P_G^{(1)},\ldots ,P_G^{(B)}.
\end{eqnarray*}
The question is how to aggregate these $B$ p-values to a
single one such that the type I error rate is still controlled. In
particular, since the $B$ p-values arise from different random splits of the
data, they are dependent, and we thus need to develop a method to aggregate
arbitrarily dependent p-values. This can be done by the following rule:
\begin{eqnarray}\label{aggreg}
& &P_G = \min\Big(1, \enspace \log(1 - \gamma_{\mathrm{min}}) \inf_{\gamma \in
    (\gamma_{\mathrm{min}},1)}  Q_G(\gamma)\Big),\nonumber\\
& &Q_G(\gamma) = q_{\gamma} \Big(\big\{P_G^{(b)}/\gamma;\ b=1,\ldots ,B
    \big\}\Big), 
\end{eqnarray}
where $q_{\gamma} \big(\{P_G^{(b)}/\gamma;\ b=1,\ldots ,B\}\big)$ is the
$\gamma$-quantile of the $B$ p-values multiplied by $1/\gamma$. The factor
$\log(1 - \gamma_{\mathrm{min}})$ guarantees to adjust for the fact that we
are searching the smallest quantiles over the range
$(\gamma_{\mathrm{min}},1)$.

\medskip
As argued for the single sample splitting procedure, the multiple sample
splitting method is valid if the screening property $\hat{S} \supseteq S_0$
holds. Thus, for asymptotic validity in terms of controlling the type I
error, we require the screening property as in \eqref{screening}. This
itself holds for the Lasso under the assumptions (A1)--(A3) discussed in
Section \ref{subsec.property}. In particular, this approach calls for a
beta-min assumption as in (A3) which is somewhat unpleasant: the p-value or
statistical test should quantify to what extent a regression parameter is
``smallish'' or ``sufficiently large'' while the beta-min assumption is simply
assuming that there are no ``smallish nonzero'' coefficients. A
  slight relaxation of the screening property is discussed in
  \citet{pbmand13}, allowing for not too many small non-zero true regression
  coefficients. 

From a computational point of view, the method requires the computational
cost $O(np \min(n,p))$ for screening the variables with the Lasso and then at
most $O(n |\hat{S}|^2)$ for inference based on the selected variables:
thus, for $p \gg n$ and since $|\hat{S}| \le n$, the total computational
cost is $O(B n^2 p)$ which is linear in the dimensionality $p$. We
typically take $B = 100$ and parallel implementation over the
  $B$ repetitions can easily be done.
  The main cost is fitting a Lasso
regression for variable screening in the setting where $p$ is very large
and $n$ is a substantial number. Computational speed-ups for the Lasso
using random projections (in sample space) have been recently proposed
\citep{pilanci2015randomized} and might be useful in practice;
similarly, computationally fast Ridge regression \citep{lu2013faster} and
thresholding \citep{shadeng11} could be used for reasonably accurate screening,
though perhaps a bit worse than Lasso \citep{pbmand13}.  

\subsubsection{Other methods}\label{subsec.despars} 
 
Other methods which do not require a beta-min assumption can be used for
statistical hypothesis testing: for a comparison, see
\citet{dezetal15}. The most prominent example is perhaps the 
de-biased or de-sparsified Lasso estimator proposed by
\citet{zhangzhang11} and further analyzed in \citet{vdgetal13}; a related
technique has been proposed in \citet{jamo13b}. A Ridge
projection method \citep{pb13} is another option, often leading to more
conservative inferential statements.

Bootstrapping the Lasso or versions of it has been proposed in
\citet{chatter11,chatter13,liuyu13} but due to the sparsity of the
underlying estimator, these approaches are exposed to the
super-efficiency phenomenon (i.e. estimation of parameters being
equal to zero is very accurate while it can be very poor for non-zero
components). Bootstrapping the de-biased Lasso estimator, where
super-efficiency does not occur, has been analyzed in
\citet{dezpbzhang16}. A very different resampling strategy for obtaining
p-values for rather general hypotheses about ``goodness of fit'' has been
proposed in \citet{shahpb15}.

Finally, one can use ``stability selection'' for obtaining statistical
error measures \citep{mebu10,shah13}: it is a very generic subsampling
technique but does not lead to rigorous p-values corresponding to the
hypothesis in \eqref{H0G} as we require it here.

\subsection{Hierarchical inference}\label{sec.hierarchical}

\paragraph{An executive summary.} Hierarchical inference is a key technique
for computationally and statistically efficient hypothesis testing and
multiple testing adjustment. It provides a convincing way to address the
main problems occurring in high-dimensional scenarios. First, due to high
pairwise absolute empirical correlation between covariates, or near linear
dependence among a small set of covariates, one cannot (or at
least not sufficiently well) identify single regression coefficients
$\beta^0_j$. However, the problem is much better posed if we ask for
identifying whether there is an association between a group of variables $G
\subseteq \{1,\ldots ,p\}$ and a response, i.e., to test a group hypothesis
as in \eqref{H0G}. Hierarchical inference is a method for sequentially
testing many such group hypotheses, thereby automatically adapting to the
``resolution level'' without the need to pre-specify the precise form or
size of the groups. 

\begin{figure}[!htb]
\begin{center}
\includegraphics[scale=0.45]{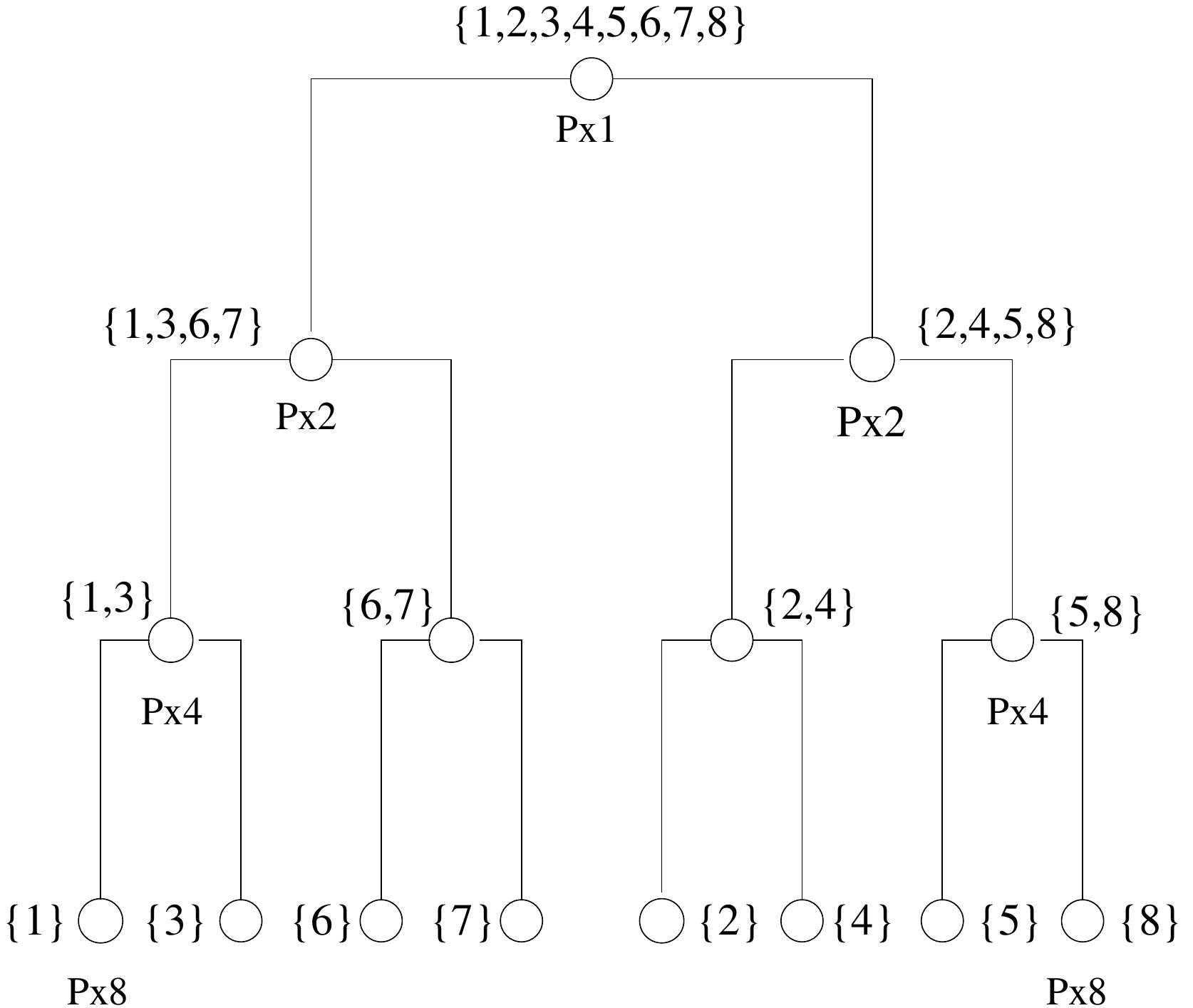}
\caption{Hierarchical grouping of 8 variables where different groups are
  denoted by $\{\ldots \}$. The capital letter ``$P$''
  is a generic notation for the raw p-value corresponding to a group
  hypothesis $H_{0,G}$ of a group $G$, which is then adjusted as in
  \eqref{B-adj2}. Since the hierarchy has 
  the same number of offspring throughout the tree, the adjustment is the
  depth-wise Bonferroni correction which amounts to 
  multiply the p-values in every depth of the tree by the number of nodes in
  the corresponding depth; no multiplicity adjustment at the top node, then
  multiplication by the factor 2 (depth 2), 4 (depth 3), and 8 (depth
ulti   4). The figure is taken from \citet{pb17}.}\label{fig1} 
\end{center}
\end{figure}

\begin{figure}[!htb]
\begin{center}
\includegraphics[scale=0.45]{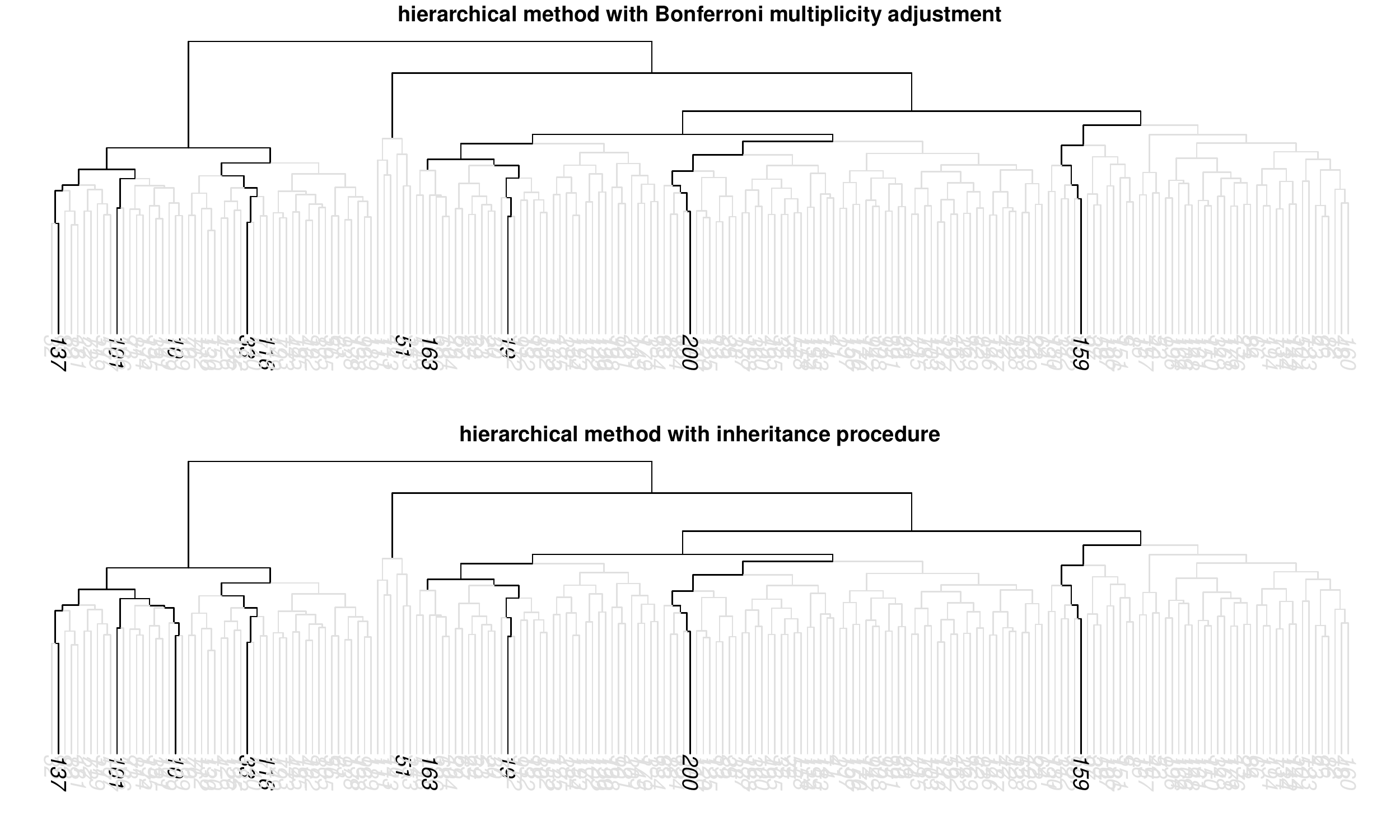}
\caption{Hierarchical inference within a tree of clustered 
    variables for a simulated example with $p = 500$ and $n = 100$. The
  numbers at the bottom in black (bold) denote the indices
    $j$ of active variables with $\beta^0_j \neq 0$ (and 
  corresponding to $H_{0,j}$ being false). The black lines graphically
    encode the significant groups of variables. Top panel: hierarchical
  procedure with the rule in \eqref{B-adj}. Bottom panel: A refined
  procedure which 
  detects in addition the single variable $10$; for 
  details see \cite{manpb16b}. The figure is taken from
  \cite{manpb16b} as well.}\label{fig2} 
\end{center}
\end{figure}

\bigskip
\noindent
The hierarchy for the inference is described in terms of a tree ${\cal T}$
where each node corresponds to a group $G (\subseteq \{1,\ldots ,p\})$ and
a group hypothesis $H_{0,G}$: the 
hierarchical constraint means that for a node (or group) $G$, any descendant
node $G'$ must satisfy $G' \subset G$. Furthermore, we require that the
child nodes of $G$ (the direct descendants of $G$) build a partition of
$G$. The tree ${\cal T}$ typically starts 
with the top node $G_{\mathrm{top}} = \{1,\ldots ,p\}$ and then branches
downward to smaller groups until the $p$
single variable nodes $\{1\},\ldots ,\{p\}$ at the bottom of the tree, see
Figures \ref{fig1} and \ref{fig2}. A
typical construction of 
such a tree is given by hierarchical clustering which results in a binary
tree, see at the end of this section. 

Given a hierarchical tree ${\cal T}$, the main idea of hierarchical
inference is to pursue testing of the groups in a sequential
fashion, starting with the top node and then successively moving down the
hierarchy until a group doesn't exhibit a significant effect. Figure
\ref{fig2} illustrates this point, showing that we might proceed rather
deep in the hierarchy at some parts of the tree whereas at other parts the
testing procedure stops due to a group which is not found to exhibit a
significant 
effect. We need some multiple testing adjustment of the p-values:
interestingly, due to the hierarchical nature, it is not overly severe at
the upper parts of the hierarchy as described below. 

The procedure works as follows. Denote by $P_G$ the raw p-value of the
statistical test for the 
null-hypothesis $H_{0,G}$ versus $H_{A,G}$ defined as in
\eqref{H0G}. We correct for multiplicity in a simple way:
\begin{eqnarray}\label{B-adj}
P_{G;\mathrm{adjusted}} = P_G \cdot p/|G|.
\end{eqnarray}
This corresponds to a depth-wise Bonferroni correction for a balanced
tree. Denote by 
$d(G)$ the level of the tree of the node (or 
group) $G$ and by $n(G)$ the number of nodes at level $d(G)$: for example,
when $G = \{1,\ldots ,p\}$ corresponds to the top 
node in a tree containing all variables, we have that $d(G) =
1$ and $n(G) = 1$. If the tree has the same number of offspring (e.g. a
binary tree with two offspring throughout the entire tree), we could also
use the unweighted version,  
\begin{eqnarray}\label{B-adj2}
\mbox{depth-wise Bonferroni correction:}\ P_{G;\mathrm{adjusted}} = P_G
    \cdot n(G),
\end{eqnarray} 
see for example \citet[eq. after eq. (22)]{pb17}. If in addition the
groups would have the same size in each level of depth (up to rounding
errors), then the 
rules in \eqref{B-adj} and \eqref{B-adj2} coincide. The formula
\eqref{B-adj2} is only given here for the sake of interpretation as a
depth-wise Bonferroni in the case of balanced trees with the same number of
off-spring. See also Figure \ref{fig1} for an illustration of such a depth-wise 
Bonferroni correction if the groups are balanced. 

The sequential nature with stopping can be formulated in terms of p-values
by adding a hierarchical constraint: 
\begin{eqnarray}\label{hier-pvalues}
P_{G;\mathrm{hierarchically-adjusted}} = \max_{G' \supset G}
  P_{G',\mathrm{adjusted}},
\end{eqnarray}
implying that once we stop rejecting a node, we cannot reject further down
in the tree hierarchy and thus, we can simply stop the procedure when a
node is not found as being significant. The main advantage of the procedure
is the statistically efficient correction for multiple testing in
\eqref{B-adj} which is much more powerful than a standard Bonferroni
correction over all the nodes in the tree, see also \eqref{B-adj2}. 
The following then holds.
\begin{prop}\citep{Meins08}\label{prop1}
Consider an arbitrary hierarchy of hypotheses tests in terms of a tree
structure ${\cal T}$. Consider the procedure described above with
depth-wise adjustment in \eqref{B-adj} and with hierarchy constraint as in
\eqref{hier-pvalues}. Then, the familywise error rate (FWER) is controlled:
that is, for $0 < \alpha < 1$, when rejecting a hypothesis $H_{0,G}$ if and
only if $P_{G;\mathrm{hierarchically-adjusted}} \le \alpha$, we have that
$\mathrm{FWER} = \PP[\mbox{at least one false rejection}] \le \alpha$.
\end{prop}

The procedure described above and justified in Proposition \ref{prop1} has
a few features to be pointed out. First, it relies on the premise 
that large groups should be easier to detect and found to be significant,
due to the fact that the identifiability is much better posed. We address
this issue at the end of this section. In fact, the
method has indeed built in 
the hierarchical constraint \eqref{hier-pvalues} that once we cannot reject
$H_{0,G}$ for 
some group $G$, we do not consider any other sub-groups of $G$ which arise
as descendants further down in the tree hierarchy. Due to the sequential
nature of the testing procedure, multiple testing adjustment for controlling the
familywise error rate is rather mild (for upper parts in the tree) as we
only correct for multiplicity at 
each depth of the tree, i.e., the root node does not
need any adjustment, and if it were found to be significant, the next
children nodes only need a correction according to the number of nodes at
depth 2 of the tree, and similarly for deeper levels; see Figure \ref{fig1}. 

Improvements over the rule in \eqref{B-adj} and \eqref{hier-pvalues} are
possible, based on exploiting the logical relationships among the tests
with the Schaffer improvement \citep{Meins08,manpb16} or using more
complete improvements from sequential testing \citep{manpb16b} based on
ideas from \citet{GoSo2010,GoFi2012}. 
Our software uses the improved hierarchical adjustment of \citet{manpb16}.
But the
essential gain in computational and statistical power is in terms of the
sequential and hierarchical nature of the procedure as illustrated in
Figures \ref{fig1} and \ref{fig2}. In particular, the method automatically
adapts to the resolution level: if the regression parameter of a single
variable is very large in absolute value, the procedure might detect such a
single variable as being significant; on the other hand, if the signal is
not sufficiently strong or if there is substantial correlation (or near linear
dependence) within a 
large number of variables in a group, the method might only identify such
a group as being significant. Figure \ref{fig2} illustrates this
point. 
Naturally, finding a large group to be significant (coarse
resolution) is much less informative than detecting a small group or even a
single variable.   

\paragraph{Methods for p-values.}
The hierarchical procedure with the rules in \eqref{B-adj} and
\eqref{hier-pvalues} requires p-values as input which are valid in the
sense that they control the type I errors of single tests. We advocate here
the use of the multi sample splitting method described in Section
\ref{subsec.multsamplspl}, implemented in our software. This method is
computationally feasible for very high dimension $p$ and it is empirically
shown to be competitive, with respect to type I error and power, over a
range of scenarios \citep{dezetal15}.   

The power of the hierarchical method is mainly hinging on the
assumption that null-hypotheses further up in the tree are easier to
reject, that is the p-values are typically getting larger when moving
downwards the tree. In low-dimensional regression problems this is
typically true when using 
partial F-tests for testing $H_{0,G}:\ \beta^0_j = 0\ \forall j \in
G$. Since our p-values rely on the partial F-test after variable screening
with the Lasso, as described in Section \ref{subsec.multsamplspl}, the same
phenomenon is expected to hold also in the high-dimensional regime. 

\paragraph{Clustering and partitioning methods for constructing the hierarchical tree.}
We describe two partitioning methods for constructing a 
  hierarchical tree of the measured SNP variables.

Motivated by the problem of identifiability among correlated variables, we
aim to construct a tree such that highly correlated variables 
are in the same groups: this can be achieved by a standard
hierarchical clustering algorithm \cite[cf.]{hartigan1975clustering}, for
example using average linkage and the dissimilarity matrix given by $1 -
(\mbox{empirical correlation})^2$. Other clustering algorithms can be used, 
for example based on canonical correlation \citep{buhl13}.

Alternatively, we can build a hierarchical tree by using the genomic
positions of SNPs. We start with an entire chromosome (or even with the
full genome sequence) and use a top-down recursive binary partitioning of
the genomic 
sequence into blocks of consecutive genomic positions, corresponding to a
binary tree, such that partitions at every depth of the tree contain about
the same number of measured SNPs. Such a spatial recursive partitioning is
computationally very fast, and it has the advantage that it can be used for
multiple studies with SNPs being measured at different locations for
different studies. We note that the approach by \citet{meijer15} also
involves spatial grouping of SNPs, using a different and computationally
more demanding procedure than hierarchical testing described above.

\subsection{Causal inference}\label{subsec.causal}

Causal inference deals with ``directional associations'', thereby going
beyond regression which is non-directional. A main tool for formalizing this
are structural equation models \citep[cf.]{pearl00}. The analogue to a
linear model in \eqref{lin.mod} is then a structural equation model with a
linear structural equation for $Y$: the data are i.i.d. realizations of 
\begin{eqnarray}\label{SEM}
& &X^{(j)} \leftarrow f_j^0\big(X^{(\pa(j))},\eps^{(j)}\big),\ j=1,\ldots ,p,\nonumber\\
& &Y \leftarrow \sum_{k \in \pa(Y)} \theta^0_{k} X^{(k)} + \eps^{(Y)},\nonumber\\
& &\eps^{(1)},\ldots ,\eps^{(p)}, \eps^{(Y)}\ \mbox{jointly independent}.
\end{eqnarray}
Here $\pa(j) = \pa_D(j)$ denotes the parental set of the node $j$ in a
graph $D$, and the
graph $D$ is assumed to be acyclic and encodes the true underlying causal
influence diagram (and the random variables $X^{(1)},\ldots ,X^{(p)},Y$ correspond
to the nodes in the graph). Furthermore, $f_j^0(.,.)$ are arbitrary
measurable potentially nonlinear functions and the ``$\leftarrow$'' symbol 
equals an algebraic ``$=$'' sign but emphasizes that the left-hand side is 
a direct ``causal'' function of the right-hand side. We note that the covariates 
are random: when conditioning on them, assuming that $Y$ is childless (see below in
Proposition \ref{prop.causal}), we have a fixed design linear model for the data
vector $Y = \bx \theta + \eps^{(Y)}$ with $\EE[\eps^{(Y)}|\bx] = 0$. 

In absence of knowing the true causal DAG $D$, the structure $D$ and the
corresponding parameter-matrix $\beta$ are typically non-identifiable from the
observational probability distribution. However, there is an interesting
exception which is relevant for the case with GWAS, namely when the node
$Y$ is childless (i.e. all edges of $Y$ point into $Y$): this simply means
that the response (e.g. disease status) is caused by the genetic SNP
biomarkers and there are no causal effects from the response to the genetic
variables. The following result holds. 
\begin{prop}\label{prop.causal}
Assume a structural equation model with a linear structural equation for
$Y$ as in \eqref{SEM} and suppose that $Y$ is childless. Consider the 
true linear regression coefficients $\beta^0$ in the 
linear regression of $Y$ versus all $X^{(1)},\ldots ,X^{(p)}$ and assume
that $\Cov((X^{(1)},\ldots
,X^{(p)})^{\top})$ is positive definite.
Then, it holds that $\beta^0_k = \theta^0_k$
for $k \in \pa(Y)$ and $\beta^0_k = 0$ for $k \notin \pa(Y)$. Thus, if
$\beta^0_k \neq 0$ it holds that 
$k \in \pa(Y)$ and there is a directed edge $X^{(k)} \to Y$ (i.e., a
direct causal effect from $X^{(k)}$ to $Y$). 
\end{prop}

\noindent
\textbf{Proof.} The DAG $D$ induces an ordering among the
variables such that $\pa_{j} \subseteq \{j-1, \ldots, 1\}$, assuming for
notational simplicity that the
variables have already been ordered (according to 
such an order). Since $Y$ is childless we can choose an ordering where $Y$
is the last element. The conditional distribution then satisfies thanks to
the Markov property:
\begin{eqnarray*}
{\cal L}\big(Y|X^{(1)},\ldots ,X^{(p)}\big) = {\cal L}\big(Y|X^{(\pa(Y))}\big).
\end{eqnarray*}
This completes the proof.\hfill$\Box$ 

\paragraph{Causal interpretation.} As a consequence, under the assumptions in Proposition \ref{prop.causal},
the inference techniques for multiple regression lead to a causal
interpretation. The main 
assumptions for such a substantially more sharpened interpretation are: (i)
the underlying true model is a structural equation model with a DAG
structure and a linear or generalized linear form for the structural
equation of $Y$ (for the latter case, using the analogous argument, we
would use a generalized linear model of $Y$ versus all $X^{(1)},\ldots
,X^{(p)}$ to 
obtain the causal variables and effects);
(ii) there are no hidden  
confounding variables between $Y$ and some of the $X^{(j)}$'s; (iii) the
response variable $Y$ is childless.
The assumption about a positive definite population covariance matrix is
weak, even in the context of GWAS; see also the discussion at the 
end of Section \ref{subsec.property}.
The last assumption (iii) is rather
plausible for GWAS since one believes that the genetic factors are the
causes for the disease and ruling out that the disease would cause a certain
constellation of genetic factors. A notable exception are retroviruses,
including e.g. HIV. The second assumption (ii) is rather
strong and perhaps the main additional assumption: relaxing it in a very
high-dimensional setting is an open problem. In view of
measuring thousands of genetic markers, the premise 
of having measured all the relevant factors is somewhat less unrealistic. The
first assumption (i) about the acyclicity of the causal influence diagram is
not important as long as there is no feedback from the response $Y$ to the $X$
variables (which is plausible for GWAS), while the requirement for a linear
or logistic form might be problematic in view of possible interactions
among the $X$-variables and/or nonlinear regression functions. The latter is a
misspecification and of the same nature as when having misspecified the
functional form in a regression model, a topic which we will discuss in
Section \ref{sec.modelmisspec}.   

One should always be careful when adopting a causal
interpretation. However, and this is a main point, the regression model
taking all the variables into account is much more appropriate than a
marginal approach where the response $Y$ is marginally regressed or
correlated to one SNP variable at a time. This has been the standard
approach over many years in GWAS, including extensions with mixed models
and adjusting for a few other covariates \citep{zhousteph14}. The approach
based on inference in a high-dimensional linear or generalized linear model
statistics comes much closer to a causal interpretation as described in
Proposition \ref{prop.causal}. And that is among the main reasons why we
believe that such multiple regression methods should lead to more
reliable results for GWAS in comparison to older marginal techniques.  

In case of complex traits, several issues with marginal testing have been
pointed out by \citet{frommlet12}. The work shows that model misspecification
can result in a severe loss  
of power to detect important SNPs and problems occur when ranking the SNPs
with respect to their p-values. Small  
correlations between causal and non-causal SNPs may lead to a 
large number of false positives.

\subsection{Misspecification of the model}\label{sec.modelmisspec}

The results in the previous sections for statistical confidence or testing
of linear model parameters rely on the correctness of a linear or
generalized linear model as in \eqref{lin.mod} or \eqref{glm.mod}.  If the
model is not correct, we have to distinguish more carefully between random
and fixed design matrix $X$ (and the latter case may also arise when
conditioning on $X$).  

For fixed design and assuming $\mathrm{rank}(X) = n$, we can always
represent any $n \times 1$ vector $f$ as $f = X \beta^*$ for some
(non-unique) $\beta^*$. Therefore, for $f = \{\EE[Y_i|X_i];i=1,\ldots ,n\}$
in a regression 
or $f = \{g(\EE[Y_i|X_i]);i=1,\ldots ,n\}$ in a generalized regression, we
can represent any (nonlinear in $x$) function $f$ evaluated at the data
points as $X \beta^*$. The only question is whether there is a
representation with a \emph{sparse} $\beta^*$. 

For random design, a fit with a linear or generalized linear model to a
potentially nonlinear model is to be interpreted as the best approximation
with a (generalized) linear model. A linear model approximation has some
interesting properties for Gaussian design but the latter is not relevant
for GWAS with discrete values for the covariates.  

A detailed treatment of model misspecification in the high-dimensional
context is given in \cite{pbvdg15}. A more general perspective in the
low-dimensional regime is given in \citet{buetal14}. 

\section{Software}\label{sec.hierinf}

The \textsf{R} package \texttt{hierinf} (available on bioconductor)
is an implementation of the
hierarchical inference described in the Section \ref{sec.hierarchical} and
it is easy to use for GWAS.  
The package is a re-implementation of the \textsf{R} package \texttt{hierGWAS} 
\citep{hierGWASpackage160} and includes new features like straightforward 
parallelization, an additional option for constructing a hierarchical 
tree based on spatially contiguous genomic positions, and the possibility
of jointly analyzing multiple datasets.  
To summarize the method, one starts by clustering the data
hierarchically. This means that the clusters can be represented by a tree.
The main idea is to pursue testing top-down and successively moving
downwards until the null-hypotheses cannot be rejected, see Section
\ref{sec.hierarchical}. The p-value of a
given cluster is calculated based on the multiple sample splitting approach
and aggregation of those p-values as described in Section
\ref{subsec.multsamplspl}. The work flow is straightforward and is
composed in two function calls. We note that the package \texttt{hierinf}
requires complete observations, i.e. no missing values in the data, because
the testing  
procedure is based on all the SNPs which is in contrast to marginal
tests. If missing  
values are present, they can be imputed prior to the analysis. This can 
be done in \textsf{R} using e.g. \texttt{mice} \citep{vanbuuren11}, 
\texttt{mi} \citep{shi11}, or \texttt{missForest} \citep{stekhoven11}.

A small simulated toy example with two chromosomes is used to demonstrate
the procedure. The toy example is taken from \citep{hierGWASpackage160} and 
was generated using \textsf{PLINK} where the SNPs were binned into different 
allele frequency ranges. The response is binary with 250 controls 
and 250 cases. 
Thus, there are $n = 500$ samples, the number of SNPs is $p = 1000$, and
there are two additional control variables  with column names ``age'' and
``sex''. The first 990 SNPs have no association  
with the response and the last 10 SNPs were simulated to have a population 
odds ratio of 2.
The functions of the package \texttt{hierinf} require 
the input of the SNP data to be a \texttt{matrix} (or a list of matrices 
for multiple datasets). We use a \texttt{matrix} instead of a 
\texttt{data.frame} since this makes computation faster. 
\begin{lstlisting}[language=R]
# load the package
library(hierinf)

# random number generator (for parallel computing)
RNGkind("L'Ecuyer-CMRG")

# We use a small build-in dataset for our toy example.
data(simGWAS) 

# The genotype, phenotype and the control variables are saved in 
# different objects.
sim.geno  <- simGWAS$x
sim.pheno <- simGWAS$y
sim.clvar <- simGWAS$clvar
\end{lstlisting}

The two following sections correspond to the two function calls in order
to perform hierarchical testing. The third section states some remarks
about running the code in parallel.  

\subsection{Software for clustering} \label{subsec.clustering}
The package \texttt{hierinf} offers two possibilities to build a 
hierarchical tree for corresponding hierarchical testing. The function 
\texttt{cluster\char`_var} performs hierarchical clustering based on 
some dissimilarity matrix and is described first. The function 
\texttt{cluster\char`_position} builds a tree based on recursive binary
partitioning of consecutive positions of the SNPs. For a short
description, see at the end of Section \ref{sec.hierarchical}. 

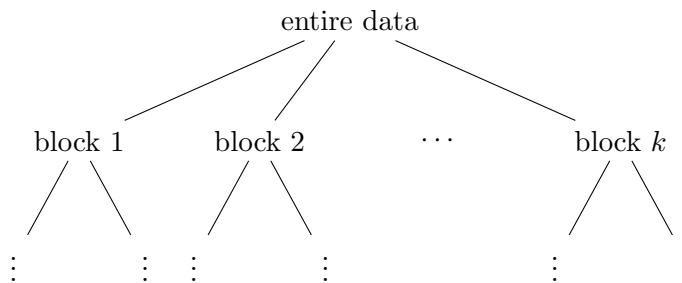
\begin{figure}[!htb]
\begin{center}
\begin{tikzpicture}[level distance=1.6cm,
  level 1/.style={sibling distance=2.4cm},
  level 2/.style={sibling distance=1.75cm}] 
\node (z) {entire data} 
  child {node (a) {block 1}
    child {node (aa) {$\vdots$}}
    child {node (ab) {$\vdots$}}
  }
  child {node (b) {block 2}
    child {node (ba) {$\vdots$}}
    child {node (bb) {$\vdots$}}
  }
  child [ missing ]
  child {node (c) {block $k$}
    child {node (ca) {$\vdots$}}
    child {node (cb) {$\vdots$}}
  };
\path (b) -- (c) node (x) [midway] {$\cdots$};
\end{tikzpicture}
\caption{The top two levels of a hierarchical tree used
to perform multiple testing. The user can optionally specify the second
level of the tree with the advantage that one can easily run the code in
parallel over different clusters in the second level, denoted by block 1,
$\ldots$, block $k$.  
A natural choice is to choose the chromosomes as the second level
of the hierarchical tree, which define a partition of the SNPs. 
If the second level is not specified, then the first split is 
estimated based on clustering the data, i.e. it is a binary split. 
The user can define the second level of the tree structure using 
the argument \texttt{block} in the functions \texttt{cluster\char`_var} / 
\texttt{cluster\char`_position}. 
The function 
\texttt{cluster\char`_var} / \texttt{cluster\char`_position} builds a separate 
binary hierarchical tree for each of the blocks.
}\label{fig3} 
\end{center}
\end{figure}

Hierarchical clustering is computationally expensive and prohibitive for
large datasets. Thus, it makes sense to pre-define dis-joint sets of SNPs
which can be clustered separately. One would typically assume that the
second level of a cluster tree structure corresponds to the blocks given
by the chromosomes as illustrated in Figure \ref{fig3}.
For the method based on binary partitioning of consecutive positions
of SNPs, we recommend to pre-define the second level of the hierarchical tree as well.
This allows to run the building of the hierarchical tree and the hierarchical
testing for each block or in our case for each chromosome in parallel, which
can be achieved using the function calls below. 
If one does not want to specify the second level of the tree, then the
argument \texttt{block} in both function calls can be omitted.

In the toy example, we define the second level of the tree structure as follows.
The first and second 500 SNPs of the SNP data \texttt{sim.geno} correspond
to chromosome 1 and chromosome 2, respectively. 
The object \texttt{block} is a \texttt{data.frame} which contains two
columns identifying the two blocks.  
The blocks are defined in the second column and the corresponding column
names of the SNPs are stored in the first column. The argument
\texttt{stringsAsFactors} of the function \texttt{data.frame} is set to
\texttt{FALSE} because we want both columns to contain integers or strings.   
  
% \newpage  
  
\begin{lstlisting}[language=R]
# Define the second level of the tree structure. 
block <- data.frame("colname" = paste0("SNP.", 1:1000), 
					"block" = rep(c("chrom 1", "chrom 2"), 
								  each = 500), 
					stringsAsFactors = FALSE)
\end{lstlisting}

% \newpage

\begin{lstlisting}[language=R]
# Cluster the SNPs
dendr <- cluster_var(x = sim.geno, 
					 block = block, 
					 # the following arguments have to be specified 
					 # for parallel computation
					 parallel = "multicore",
					 ncpus = 2)
\end{lstlisting}

By default, the function \texttt{cluster\char`_var} uses the agglomeration
method average linkage and the  
dissimilarity matrix given by $1 - (\mbox{empirical correlation})^2$.

Alternatively, \texttt{cluster\char`_position} builds a hierarchical tree
using recursive binary partitioning of consecutive genomic positions of the
SNPs. 
As for \texttt{cluster\char`_var}, the function 
can be run in parallel if the argument block defines the second level
of the hierarchical tree.

\begin{lstlisting}[language=R]

# Store the positions of the SNPs.
position <- data.frame("colnames" = paste0("SNP.", 1:1000), 
                       "position" = seq(from = 1, to = 1000), 
                       stringsAsFactors = FALSE)
                       

# Build the hierarchical tree based on the position.
# The argument block defines the second level of the tree structure. 
dendr.pos <- cluster_position(position = position, 
							  block = block,
							  # the following arguments have to be 
							  # specified for parallel computation
							  parallel = "multicore",
							  ncpus = 2)
\end{lstlisting}

\subsection{Software for hierarchical testing} \label{subsec.hiertesting}

The function \texttt{test\char`_hierarchy} is executed after the function 
\texttt{cluster\char`_var} or \texttt{cluster\char`_position} since it requires 
the output of one of those two functions as an input (argument \texttt{dendr}).

The function \texttt{test\char`_hierarchy} first randomly splits the data
into two halves (with respect to the observations), by default \texttt{B = 50} 
times, and performs variable screening on the second half. Then, the function 
\texttt{test\char`_hierarchy} uses those splits and corresponding selected 
variables to perform the hierarchical testing according to the tree defined 
by the output of one of the two functions \texttt{cluster\char`_var} or 
\texttt{cluster\char`_position}. 

As mentioned in Section \ref{subsec.clustering}, we can exploit the
proposed hierarchical structure which assumes the chromosomes to form the
second level of the tree structure as illustrated in Figure
\ref{fig3}. This allows to run the testing in parallel for each 
block, which are the chromosomes in the toy example.  

The following function call performs first the global null-hypothesis test
for the group containing all the variables/SNPs and continues
testing in the hierarchy of the two chromosomes and their children.  
\begin{lstlisting}[language=R]
# Test the hierarchy using multi sample split
set.seed(1234)
result <- test_hierarchy(x = sim.geno, 
						 y = sim.pheno, 
						 clvar = sim.clvar,
						 # alternatively: dendr = dendr.pos 
						 dendr = dendr,
						 family = "binomial",
						 # the following arguments have to be 
						 # specified for parallel computation
						 parallel = "multicore",
						 ncpus = 2)				 
\end{lstlisting}

The function \texttt{test\char`_hierarchy} allows to fit models with
continuous or binary response, the latter being based on logistic
regression. The argument \texttt{family} is set to \texttt{"binomial"}
because the response variable in the toy example is binary. 

The output looks as follows:
\begin{lstlisting}[language=R]
> print(result, n.terms = 4)
  block   p.value   significant.cluster                        
1 chrom 1 NA        NA                                         
2 chrom 2 0.0489170 SNP.605, SNP.792, SNP.636, SNP.857, ... [8]
3 chrom 2 0.0020718 SNP.992                                    
4 chrom 2 0.0001637 SNP.991                                    
5 chrom 2 0.0047820 SNP.1000                                   
6 chrom 2 0.0065060 SNP.994                                    
7 chrom 2 3.858e-05 SNP.993 
\end{lstlisting}

The output shows significant groups of SNPs or even single SNPs if there is
sufficiently strong signal in the data. The block names, the p-values, and
the column names (of the SNP data) of the significant clusters are
returned. There is no significant cluster in chromosome 1. 
That's the reason why the p-value and the column names of the significant 
cluster are \texttt{NA} in the first row of the output.
Note that the large significant cluster in the second row of the output is 
shortened to better fit on screen. In our toy example, the last 8 column names 
are replaced by ``\texttt{... [8]}''. The maximum number of terms can be 
changed by the argument \texttt{n.terms} of the \texttt{print} function. 
One can evaluate the object \texttt{result} in the console and the default 
values of the \texttt{print} function are used. In this case, it would only 
display the first 5 terms.

The only difference in the \textsf{R} code 
when using a hierarchical tree based on binary recursive partitioning of the
genomic positions of the SNPs (whose output is denoted as \texttt{dendr.pos}) 
is to specify the corresponding hierarchy: 
\texttt{test\char`_hierarchy(..., dendr = dendr.pos, ...).}

 We can access part of the output 
by \texttt{result\$res.hierarchy} which we use below to calculate the 
$\mbox{R}^2$ value of the second row of the output, i.e. 
\texttt{result\$res.hierarchy[[2, "significant.cluster"]]}. Note that we 
need the double square brackets to access the column names stored in the column 
\texttt{significant.cluster} of the output since the last column is a list 
where each element contains a character vector of the column names. The two
other columns containing the block names and the p-values can both be indexed 
using single square brackets as for any \texttt{data.frame}, e.g. 
\texttt{result\$res.hierarchy[2, "p.value"]}.

\begin{lstlisting}[language=R]
> (coln.cluster <- result$res.hierarchy[[2, "significant.cluster"]])
[1] "SNP.605" "SNP.792" "SNP.636" "SNP.857" "SNP.858" "SNP.911"
[7] "SNP.571" "SNP.998" "SNP.708" "SNP.867" "SNP.612" "SNP.932"
\end{lstlisting} 

The function \texttt{compute\char`_r2} calculates the adjusted $\mbox{R}^2$ value or
coefficient of determination of a cluster for a continuous response. The
Nagelkerke's $\mbox{R}^2$ \citep{nagelkerke91} is calculated for a binary
response as e.g. in our toy example.

\begin{lstlisting}[language=R]
> compute_r2(x = sim.geno, y = sim.pheno, clvar = sim.clvar, 
             res.test.hierarchy = result, family = "binomial", 
             colnames.cluster = coln.cluster)
[1] 0.06324339
\end{lstlisting}  
The function \texttt{compute\char`_r2} is based on multi-sample splitting. 
The $\mbox{R}^2$ value is calculated per split based on the second half of 
observations and based on the intersection of the selected variables and  
the user-specified cluster. Then, the $\mbox{R}^2$ values are
averaged over the different splits.
If one does not specify the argument \texttt{colnames.cluster}, then the
$\mbox{R}^2$ value of the whole dataset is calculated.

\subsection{Software for parallel computing} \label{subsec.parallel} 

The function calls of \texttt{cluster\char`_var}, \texttt{cluster\char`_position}, 
and \texttt{test\char`_hierarchy} above are evaluated in parallel since we set 
the arguments \texttt{parallel = "multicore"} and \texttt{ncpus = 2}. 
The argument \texttt{parallel} can be set to \texttt{"no"} for serial 
evaluation (default value), to \texttt{"multicore"} for parallel evaluation 
using forking, or to \texttt{"snow"} for parallel evaluation using a parallel 
socket cluster (PSOCKET); see below for more details. The argument 
\texttt{ncpus} corresponds to the number of cores to be used for parallel 
computing. We use the \texttt{parallel} package for our implementation 
which is already included in the base \textsf{R} installation 
\citep{rcoreteam17}. 

The user has to select the ``L'Ecuyer-CMRG'' pseudo-random number 
generator and set a seed such that the parallel computing of \texttt{hierinf} 
is reproducible. This pseudo-random number generator can be selected by 
\texttt{RNGkind("L'Ecuyer-CMRG")} and has to be executed once for every new  
\textsf{R} session; see \textsf{R} code at the beginning of Section 
\ref{sec.hierinf}. This allows us to create multiple streams of pseudo-random 
numbers, one for each processor / computing node, using the \texttt{parallel} 
package; for more details see the vignette of the \texttt{parallel} package 
published by \citet{rcoreteam17}.

We recommend to set the argument \texttt{parallel = "multicore"} which will 
work on Unix/Mac (but not Windows) operation systems. The function is 
then evaluated in parallel using forking which is leaner on the memory usage.  
This is a neat feature for GWAS since e.g. a large SNP dataset does not
have to be copied to the new environment of each of the processors. Note 
that this is only possible on a multicore machine and not on a cluster. 

On all operation systems, it is possible to create a parallel socket 
cluster (PSOCKET) which corresponds to setting the argument 
\texttt{parallel = "snow"}. This means that the computing nodes or 
processors do not share the memory, i.e. an \textsf{R} session with an 
empty environment is initialized for each of the computing nodes or processors. 

How many processors should one use? If the user specifies the second level 
of the tree, i.e. defines the \texttt{block} argument of the functions 
\texttt{cluster\char`_var} / \texttt{cluster\char`_position} and 
\texttt{test\char`_hierarchy}, then the building of the hierarchical tree and 
the hierarchical testing can be easily performed in parallel across the different 
blocks. Note that the package can make use of as many processors as there are 
blocks, say, 22 chromosomes. In addition, the multi sample splitting 
and screening step, which is performed inside the function 
\texttt{test\char`_hierarchy}, can always be executed in parallel regardless 
if we defined blocks or not. It can make use of at most $B$ processors where 
$B$ is the number of sample splits. 

\subsection{Illustration: hierarchical inference on real datasets}\label{subsec.hierinfrealdatasets} 

Hierarchical inference for GWAS has been successfully applied in some of
our own previous work \citep{buzduganetal16,klasenetal16}.  

One dataset is about type 1 diabetes with a binary response
  variable (``healthy''/``diseased''): \citet{Well1} measured  
  500'568 SNPs of 2'000 cases and 3'000 controls. Some of the 
  results from \citet{buzduganetal16} are described in Table \ref{tab1}.
\citet{buzduganetal16} found a
significant association of the response and eight single SNPs: five of those SNPs
have been found to be significant in the study of \citet{Well1}. One of the
other three SNPs was found to have a moderate association in an
independent study \citep{plagnol11}.  

\citet{buzduganetal16} identified two small significant groups of SNPs for 
the type 2 diabetes dataset which has the same sample size and number of 
SNPs as the type 1 diabetes dataset. Their results are described in Table 
\ref{tab2}. Both groups contain one SNP which was originally found significant 
by \citet{Well1}. There are two SNPs, one in each 
of the two groups, that were shown significant in an independent study by \citet{zeggini07} 
and only one of those two SNPs by \citet{scott07}.  

\renewcommand{\arraystretch}{1.2}

\begin{table}[!htb]
\begin{center}
  \begin{tabular}{l*{3}{r}l}
  \hline   
  Significant group of SNPs & Chr & Gene & p-value & $\mbox{R}^2$  \\
  \hline
  rs6679677  & 1  & PHTF1      & $3.6 * 10^{-11}$ & 0.03  \\
  rs17388568 & 4  & ADAD1      & $2.7 * 10^{-2}$  & 0.006 \\
  rs9272346  & 6  & HLA-DQA1   & $2.4 * 10^{-3}$  & 0.17  \\
  rs9272723  & 6  & HLA-DQA1   & $2.2 * 10^{-4}$  & 0.17  \\
  rs2523691  & 6  & intergenic & $6.04 * 10^{-5}$ & 0.004 \\
  rs11171739 & 12 & intergenic & $1.3 * 10^{-2}$  & 0.01  \\
  rs17696736 & 12 & NAA25      & $6.5 * 10^{-4}$  & 0.018 \\
  rs12924729 & 16 & CLEC16A    & $3.4 * 10^{-2}$  & 0.007 \\
  \hline
  \end{tabular}
  \caption{List of small significant groups of SNPs for type 1 diabetes. 
    The smallest groups of SNPs whose null hypothesis was rejected are displayed. 
    The SNPs in this group are jointly significant. The rsIDs are taken from 
    dbSNP. Chromosome is abbreviated by Chr. If the group of SNPs belongs to 
    a gene, then the gene symbol from Entrez Gene is stated in the corresponding 
    column. The p-values are adjusted for multiple testing (controlling the FWER) 
    and the $\mbox{R}^2$ value is the explained variance by the group of SNPs. 
  	The table is taken from \citet{buzduganetal16}.} \label{tab1} 
\end{center}
\end{table}

\begin{table}[!htb]
\begin{center}
  \begin{tabular}{l*{3}{r}l}
  \hline   
  Significant group of SNPs & Chr & Gene & p-value & $\mbox{R}^2$  \\
  \hline
  rs4074720, rs10787472, rs7077039, rs11196208, rs11196205, 	& 10  	& TCF7L2    & $1.7 * 10^{-5}$ 	& 0.015 \\
  \quad rs10885409, rs12243326, rs4132670, rs7901695, rs4506565 &   	&   		&  					&   	\\[1.5mm]
  rs9926289, rs7193144, rs8050136, rs9939609  					& 16  	& FTO      	& $4.7 * 10^{-2}$  	& 0.007 \\
  \hline
  \end{tabular}
  \caption{List of small significant groups of SNPs for type 2 diabetes. 
  The smallest groups of SNPs whose null hypothesis was rejected are displayed. 
  The SNPs in this group are jointly significant. The rsIDs are taken from 
  dbSNP. Chromosome is abbreviated by Chr. If the group of SNPs belongs to 
  a gene, then the gene symbol from Entrez Gene is stated in the corresponding 
  column. The p-values are adjusted for multiple testing (controlling the FWER) 
  and the $\mbox{R}^2$ value is the explained variance by the group of SNPs. 
  The table is taken from \citet{buzduganetal16}.} \label{tab2} 
\end{center}
\end{table}

\citet{klasenetal16} compare hierarchical testing with linear mixed effect
models and stress that the hierarchical testing seems less exposed to
population structure and often does not need a corresponding correction. 
One of the studied datasets is about the association between the root 
development and the genotype of 201 world-wide collected natural Arabidopsis accessions.
They found one significant locus with a linear mixed effect model whereas 
with the hierarchical testing they discovered three additional loci which 
are located in two neighboring genes.  
\citet{klasenetal16} made a follow-up randomized treatment-control
experiment to validate an effect of one of these two genes on the root
growth (namely the PEPR2 gene): it turned out to be successful exhibiting a
significant effect. 

\section{Meta-analysis for several datasets} \label{sec.meta}

Consider the general situation with $m$ datasets 
\begin{eqnarray*}
Y^{(\ell)},\bx^{(\ell)},\ \ell=1,\ldots ,m,
\end{eqnarray*}
with $n_{\ell} \times 1$ response vector $Y^{(\ell)}$ and $n_{\ell} \times
p_{\ell}$ design matrix $\bx^{(\ell)}$. For each of them we assume a
potentially high-dimensional linear model
\begin{eqnarray*}
Y^{(\ell)} = \bx^{(\ell)} \beta^{(\ell)} + \eps^{(\ell)},
\end{eqnarray*}
with $\eps^{(\ell)}_1,\ldots , \eps^{(\ell)}_{n_{\ell}}$ i.i.d. having
$\EE[\eps^{(\ell)}_i] = 0,\ \Var(\eps^{(\ell)}_i) =
(\sigma^{(\ell)})^2$. To simplify notation, we drop here the superscript
``$^0$'' for denoting the true underlying parameter. 
Note that the treatment for generalized linear models is analogous. 

For simplicity, we consider here only the case where the measured covariates
are the same across all the $m$ datasets. This implies that $p_{\ell} \equiv
p$ for all $\ell=1,\ldots ,m$. We consider the null-hypothesis for single
variables 
\begin{eqnarray}\label{nullhypj}
\tilde{H}_{0,j}:\ \beta^{(\ell)}_j = 0\ \mbox{for all}\ \ell = 1,\ldots ,m,
\end{eqnarray}
versus the alternative 
\begin{eqnarray}\label{alterj}
\tilde{H}_{A,j}:\ \mbox{there exists}\ \ell \in \{1,\ldots ,m\}\
  \mbox{with}\ \beta^{(\ell)}_{j} \neq 0.
\end{eqnarray}
For groups of variables $G \subseteq \{1,\ldots ,p\}$ we have the analogous
hypotheses:
\begin{eqnarray}\label{nullhypG}
\tilde{H}_{0,G}:\ \beta^{(\ell)}_G \equiv 0\ \mbox{for all}\ \ell = 1,\ldots ,m,
\end{eqnarray}
versus the alternative 
\begin{eqnarray}\label{alterG}
\tilde{H}_{A,G}:\ \mbox{there exists}\ j \in G\ \mbox{and}\ \ell \in \{1,\ldots ,m\}\
  \mbox{with}\ \beta^{(\ell)}_{j} \neq 0.
\end{eqnarray}
If $\tilde{H}_{0,j}$ is rejected we conclude that covariate $j$ is
significant in at least one dataset. 
From an abstract point of view, $\tilde{H}_{0,j}$ or $\tilde{H}_{0,G}$ as
in \eqref{nullhypj} or \eqref{nullhypG} are again group hypothesis with
coefficient indices in the 
group $(\ell,j) \in \{1,\ldots , m\} \times \{j\}$ or $(\ell,j) \in
\{1,\ldots , m\} \times G$, respectively. 

A simple way to test the hypotheses in \eqref{nullhypj} or \eqref{nullhypG}
is to aggregate the corresponding p-values for the datasets $\ell=1,\ldots
,m$. 
Denote by
$P_{G}^{(\ell)}$ the 
p-value for testing the null-hypothesis $H_{0,G}^{(\ell)}:\
\beta_G^{(\ell)} \equiv 0$ for the dataset $\ell$.

We advocate here the use of Tippett's rule \citep{tippett31}: 
\begin{eqnarray} \label{aggreg-Tippett}
& &P_{\mathrm{Tippett};G} \, = \, 1- \big(1 -  \min\{P_{G}^{(\ell)},\ \ell=1,\ldots ,m\}\big)^m, % \\
\end{eqnarray}
where $P_{G}^{(1)},\ldots, P_{G}^{(m)}$ are the raw p-values. This aggregated
p-value 
controls the familywise error rate at level $\alpha$ for the decision 
rule: reject $\tilde{H}_{0,G}$ 
if and only if $P_{\mathrm{Tippett};G} \le
  \alpha$ for some significance level $\alpha$.

Alternatively, p-values can be aggregated by Stouffer's rule
\citep{stouffer49}:  
\begin{eqnarray}  \label{aggreg-Stouffer}
P_{\mathrm{Stouffer};G} \, = \, \Phi\Big(\sum_{\ell=1}^m w_{\ell}
  \Phi^{-1}\big(P_{G}^{(\ell)}\big)\Big),\ w_{\ell} = \sqrt{n_{\ell}/n},\ n = \sum_{\ell=1}^m
  n_{\ell}.  
\end{eqnarray}
This p-value controls the familywise error rate at level $\alpha$ for the decision
rule: reject $\tilde{H}_{0,G}$ if and only if
$P_{\mathrm{Stouffer};G} \le \alpha$. 

\begin{figure}%
    \centering
    \subfigure[Tippett's rule.]{\includegraphics[width = 0.45\textwidth]{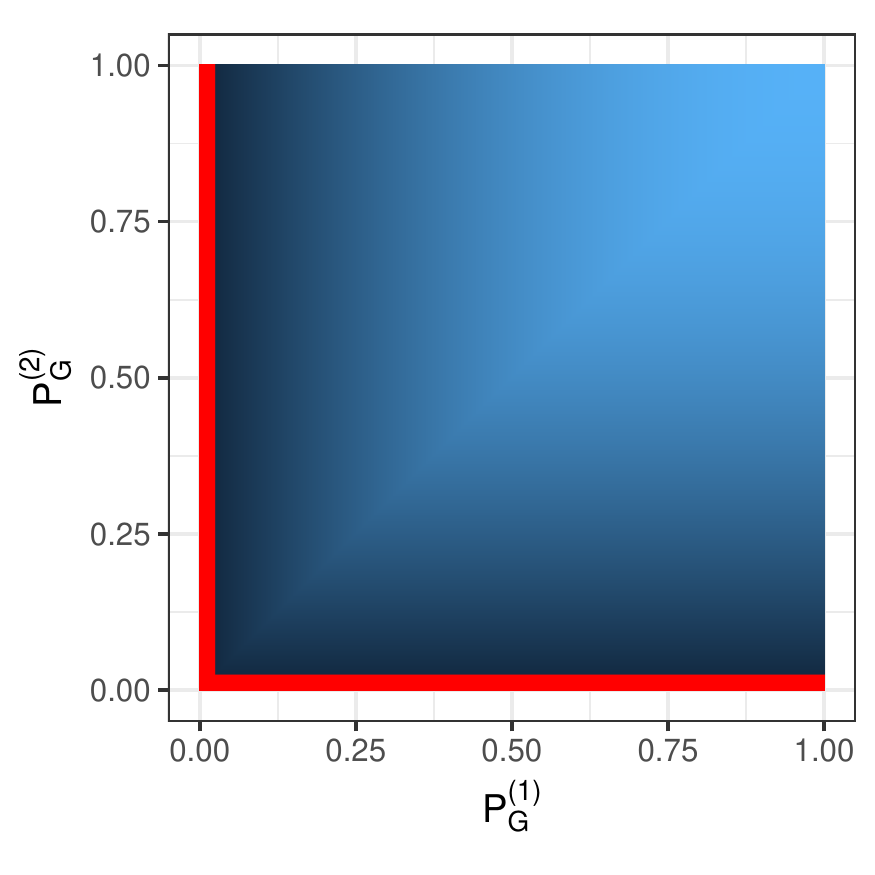}}%
    \qquad
	\subfigure[Stouffer's rule.]{\includegraphics[width = 0.45\textwidth]{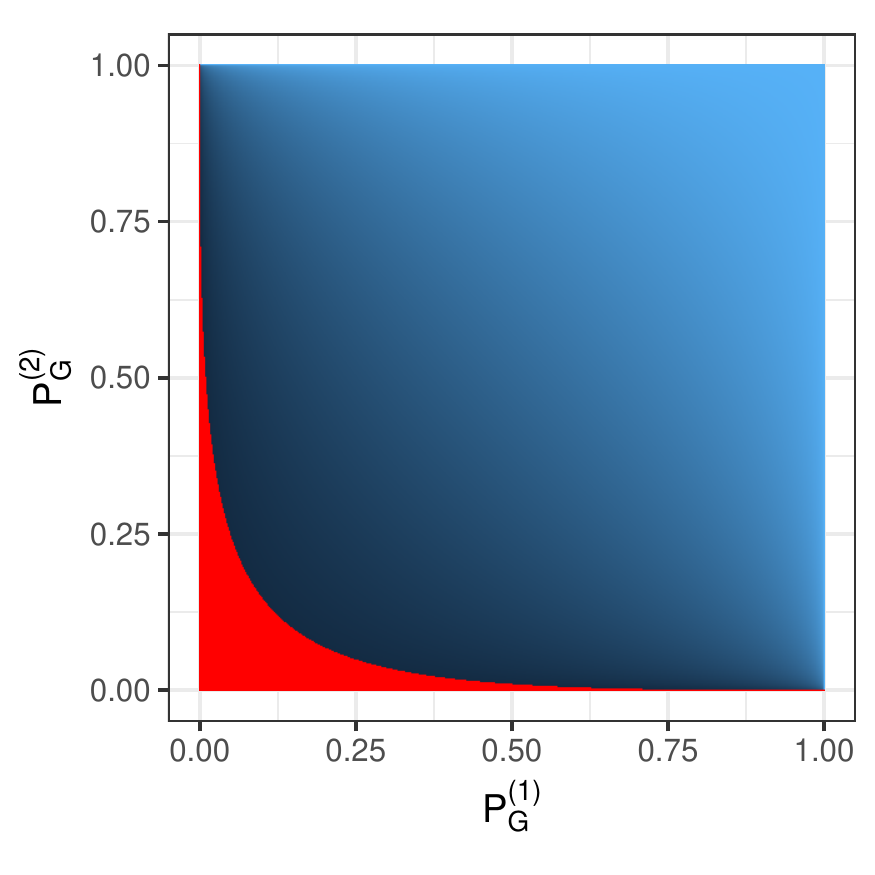}}%
    \caption{Aggregated p-value based on two datasets. The red areas highlight the aggregated 
    p-values which are below 0.05.}%
    \label{fig.agg}%
\end{figure}

For illustration purposes, we consider the case $m = 2$ in Figure
\ref{fig.agg}. The individual p-values $P_{G}^{(1)}$ and $P_{G}^{(2)}$ 
are plotted on the $x$- and $y$-axis, respectively and the aggregated
values $P_{\mathrm{Tippett};G}$ and 
$P_{\mathrm{Stouffer};G}$ are color-coded in the respective plots. Both red 
areas are equal to 0.05. The difference between the two plots is 
that Stouffer's rule is more powerful in the case of two datasets with weak 
signal and Tippett's rule is more powerful in the case of one dataset with a 
strong signal and the other having a very weak or no signal. 

We advocate the use of Tippett's rule because it performs best in our 
simulations for all scenarios; see Figure \ref{fig4} and Section 
\ref{subsec.empiricalmeta} for more details. This seems
partially due to the hierarchical multiple sample splitting inference
method which is unstable, especially for weaker signals: it happens 
fairly often that a cluster turns out to be clearly significant in 
one dataset and not significant at all in another, a situation where 
Tippett's rule is much more powerful. See also the paragraph at the 
end of Section   \ref{subsec.empiricalmeta.m10}

The naive (and conceptually wrong) approach would be to pool the different
datasets and proceed as if it would be one homogeneous dataset. This would
then result in p-values $P_{\mathrm{pooled};G}$ by using the methods from
Section \ref{sec.highdiminfer}. 

\paragraph{Fast computational methods for pooled GWAS.} There has been a
considerable interest for fast algorithms for GWAS with very large sample
size in the order of $10^5$; see \citet{lippert11, zhousteph14}.
Often though, such large sample size comes from pooling different studies 
or sub-populations. We argue in favor of meta analysis and aggregating 
corresponding p-values. Besides more statistical robustness against 
heterogeneity (arising from the different sub-populations), meta-analysis 
is also computationally very attractive: the computations can be trivially 
implemented in parallel for every sub-population and the p-value aggregation 
step comes essentially without any computational cost.

\subsection{Empirical results for aggregating p-values and pooling of two
  datasets}\label{subsec.empiricalmeta} 
We perform a simulation study to compare power and error rate for three
methods. We consider aggregating the p-values using Tippett's rule
as described in \eqref{aggreg-Tippett}, Stouffer's method in
\eqref{aggreg-Stouffer}, and pooling the datasets. The latter is a very  
simple method where we ignore that we deal with different datasets or
studies and run the hierarchical testing on the pooled set of observations
but allowing for a different intercept per dataset.  

For simplicity, we consider the case of two datasets, i.e. $m = 2$. Denote
the true underlying parameter by $\beta^{(\ell)}$ for $\ell = 1, 2$ and
the corresponding active set by  
\begin{equation*}
S_0^{(\ell)} = \big\{j; \ \beta_j^{(\ell)} \neq 0 \big\}, \ \ell = 1, 2.
\end{equation*}
As an easy case we assume here that the active sets of the two datasets
coincide $S_0^{(1)} = S_0^{(2)}$ and 
that the true underlying parameters $\beta^{(1)}$ and $\beta^{(2)}$
take the values $1$ and $-1$ on the active set, respectively. If one pools
the two datasets, then those effects roughly cancel each other (when the
datasets have approximately the same sample sizes). On the other hand, when
aggregating p-values from individual datasets, effects do not cancel out. 

To compare the two methods, we generate semi-synthetic data which is based
on data from openSNP (\url{https://opensnp.org/}), where people donate
their raw genotypic data into the public domain (using CC0 license). 
We generate two datasets $\bx^{(\ell)}$, $\ell = 1, 2$, with $n = 300$
observations each and two (consecutive) blocks of $500$ SNPs from
chromosome 1 and 2, respectively. This makes in total $p = 1000$ SNPs.  
Both datasets share the same 1000 SNPs and are kept fixed 
for the simulation.

For the generation of those two datasets, columns with many missing 
values are excluded and remaining columns are imputed using the median.
We further exclude columns with 
standard deviation zero and omit columns in order not to have a set 
of collinear columns of set size up to 10. 

For each simulation run, we randomly pick an active set of size 10 which 
is the same for both datasets. Thus, $S_0 = S_0^{(1)} = S_0^{(2)}$. 
We simulate a continuous response using 
\begin{equation*}
Y^{(\ell)} = \bx^{(\ell)} \beta^{(\ell)} + \eps^{(\ell)}, \ \ell = 1, 2,
\end{equation*}  
where each element of $\eps^{(\ell)}$ is drawn from a 
$\mathcal{N}\big(0, (\sigma^{(\ell)})^2\big)$-distribution. 
For the simulation, we vary the values of $(\sigma^{(2)})^2$ and the 
values of $\beta^{(\ell)}$ for the corresponding elements which are 
in the active set $S_0^{(\ell)}$, $\ell = 1, 2$. 

The two datasets play different roles. The variance $(\sigma^{(1)})^2 = 1$ 
is fixed for the dataset $Y^{(1)}, \bx^{(1)}$ and only the value of the 
non-zero elements of $\beta^{(1)}$ are varied. The dataset 
$Y^{(1)}, \bx^{(1)}$ carries a strong signal in general. 
The dataset $Y^{(2)}, \bx^{(2)}$ shows a weak signal especially when we
inflate the variance $(\sigma^{(2)})^2$. The elements of
$\beta^{(2)}$ corresponding to the active set take only values $0$,
$0.5$ and $1$. 

We use a modified definition of the power as the performance measure for
the simulation study because it takes the size of the significant clusters
into account. We define the adaptive power by 
\begin{equation*}
\mbox{Power}_{\mbox{\footnotesize{adap}}} = \frac{1}{|S_0|} \sum\limits_{C \, \in \, \mbox{\footnotesize{MTD}}} \frac{1}{|C|}
\end{equation*} 
where MTD stands for Minimal True Detections which means that the cluster
has to be significant (``Detection''), there is no significant subcluster
(``Minimal''), and the cluster contains at least one active variable
(``True''). This is the same definition as in \cite{manpb16}.  

\begin{figure}[!htb]
\begin{center}
\includegraphics[scale=0.8]{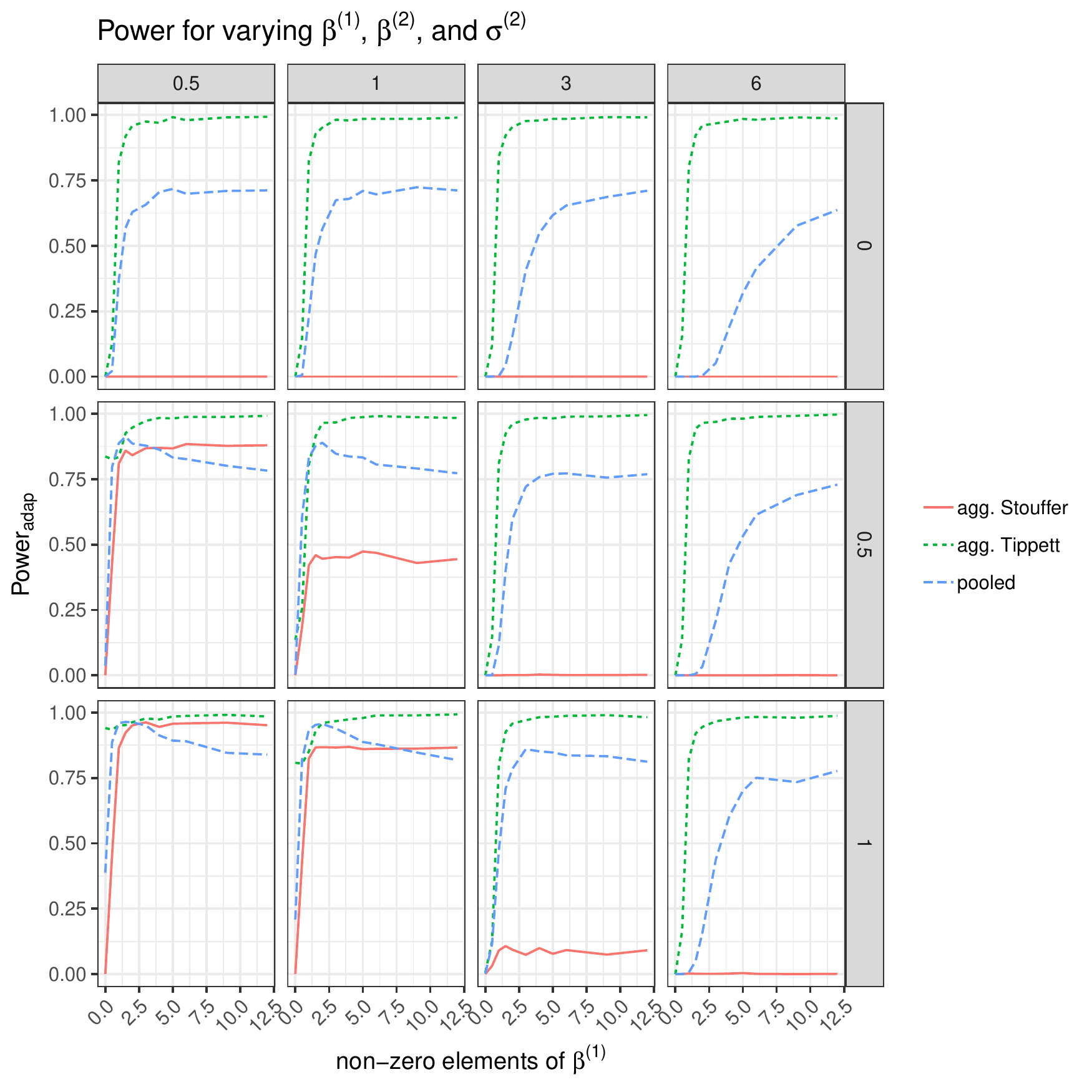}
\caption{Two datasets: comparison of the adaptive power of aggregating the p-values using
  Tippett's rule, Stouffer's rule or by simply 
pooling multiple studies. The 
values of the active or non-zero element of both datasets are varied, i.e. 
$\beta^{(1)} \in \{0, 0.5, 1, 1.5, 2, 3, 4, 5, 6, 9, 12\}$ ($x$-axis) and 
$\beta^{(2)} \in \{0, 0.5, 1\}$ (multi panels: rows). The standard deviation 
of the error is varied for the second dataset, i.e. $\sigma^{(1)} = 1$ and
$\sigma^{(2)} \in \{0.5, 1, 3, 6\}$ (multi panels: columns). 
The active set is of size 10 and is randomly selected for each simulation
run. The adaptive power was calculated based on 100 independent simulations
for each combination of the parameters.}\label{fig4}   
\end{center}
\end{figure}

\begin{figure}[!htb]
\begin{center}
\includegraphics[scale=0.8]{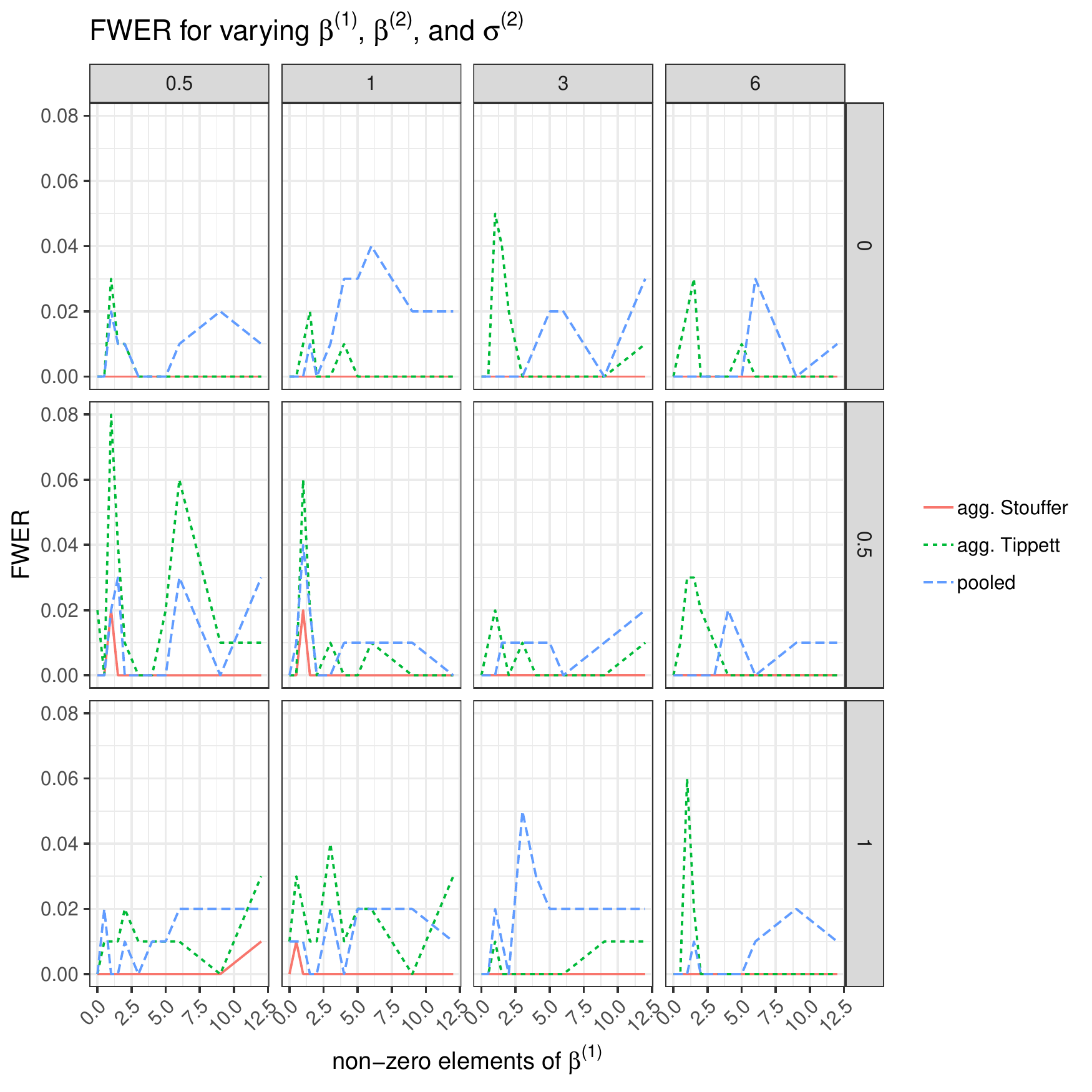}
\caption{Two datasets: comparison of the familywise error rate (FWER) of aggregating the 
p-values using Tippett's rule, Stouffer's rule or by simply pooling
multiple studies. All three methods control the FWER at level 0.05. 
}\label{fig6}  
\end{center}
\end{figure}

Figure \ref{fig4} illustrates the adaptive power of the
simulation study. 
Aggregating the p-values using Tippett's rule is clearly 
better than pooling and outperforms Stouffer's method. The two
different aggregation methods and  
their advantages have been already discussed at the beginning of Section
\ref{sec.meta}. Pooling the datasets seems to work fine especially for
situations  
where the values of the non-zero elements of $\beta^{(1)}$ and 
$\beta^{(2)}$ are similar and the standard deviation 
$\sigma^{(2)}$ takes values 0.5 or 1, i.e. similar standard deviations 
for both datasets. 
But in these situations, aggregating the p-values using Tippett's rule 
works comparably well. We note that with pooling, the power can
  slightly decrease when the true regression parameters in one dataset
  increase in size: this is somewhat counter-intuitive but might occur
  because misspecification with pooling can become stronger when increasing
  the regression parameters in one dataset.
In general aggregation with Tippett's rule 
performs more reliably than pooling since the latter is
conceptually wrong.
Figure \ref{fig6} illustrates that the familywise error
rate (FWER) is controlled for all three methods, for most scenarios even 
conservatively.

The conceptual correctness together with the results of the simulation
study support our recommendation to aggregate the p-values from different
datasets or studies rather than a simple-minded pooling of the datasets. 
Aggregating the p-values of multiple studies is very easy to perform
using the \textsf{R} package \texttt{hierinf} as described in Section 
\ref{subsec.multiplestudies}. 

\subsection{Empirical results for aggregating p-values and pooling of multiple datasets}\label{subsec.empiricalmeta.m10}

We consider two simulations for the case of $m = 10$ semi-synthetic datasets 
$Y^{(\ell)}$, $\bx^{(\ell)}$, 
$\ell = 1, \ldots, 10$, with $n = 150$ observations and $p = 10000$ SNPs 
as described in Section \ref{subsec.empiricalmeta}. The response is
simulated as 
\begin{equation*}
Y^{(\ell)} = \bx^{(\ell)} \beta^{(\ell)} + \eps^{(\ell)}, \ \ell = 1,
\ldots, 10, 
\end{equation*}  
where each element of $\eps^{(\ell)}$ is drawn from a 
$\mathcal{N}(0, 1)$-distribution, i.e. all the variances are kept fixed. 

We examine two scenarios where the support of the parameter vectors
  $\beta^{(\ell)}$ is the same across all datasets. In particular, the
  non-zero elements of $\beta^{(\ell)}$, $\ell = 1, \ldots, 5$,
  respectively, $\ell = 1, \ldots, 8$,  
are varied by one number while the non-zero elements of $\beta^{(k)}$ 
of the remaining datasets are equal to 0.5. 

\begin{figure}%
    \centering
    \subfigure[Adaptive Power.]{\includegraphics[width = 0.45\textwidth]{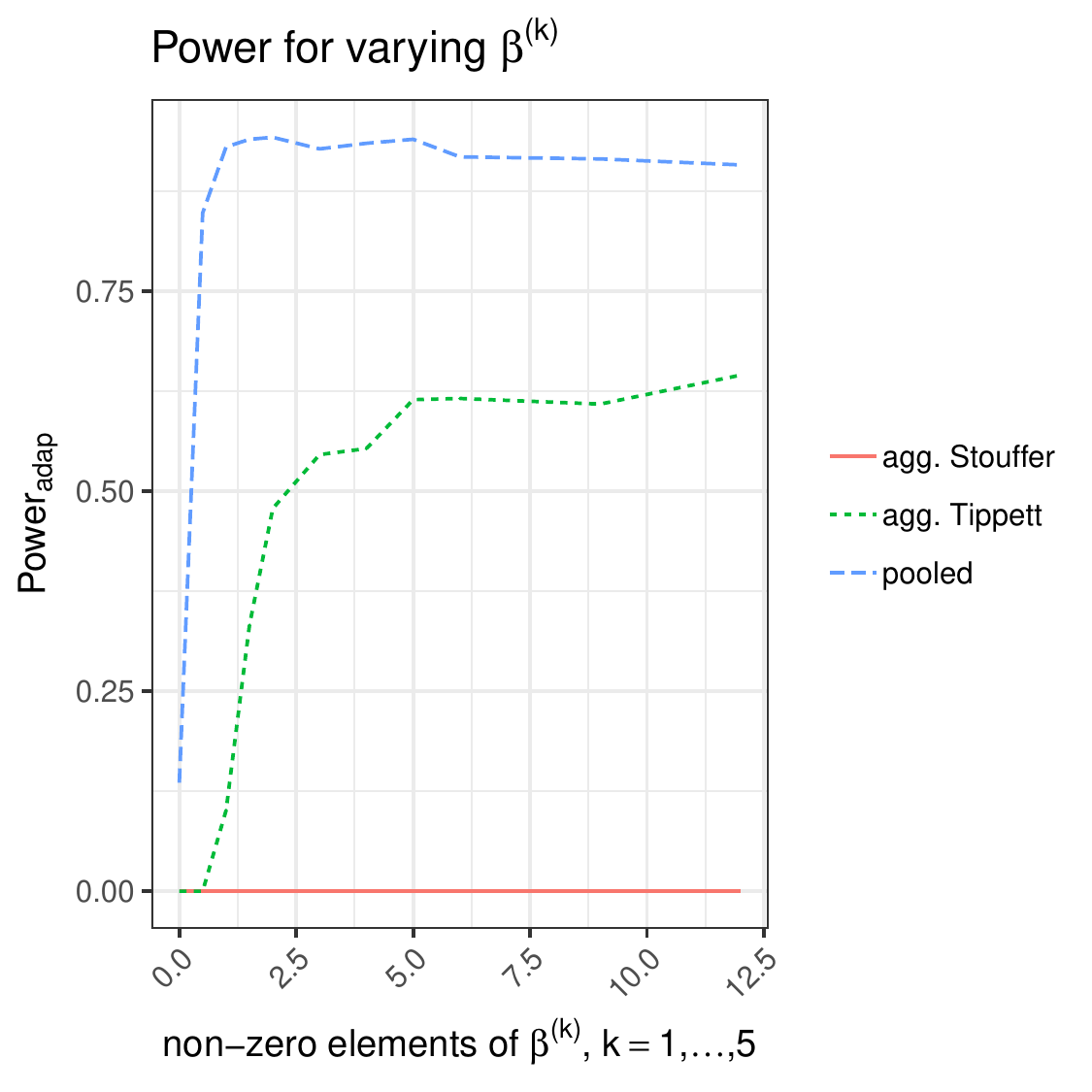}}%
    \qquad
	\subfigure[Family-Wise Error Rate.]{\includegraphics[width = 0.45\textwidth]{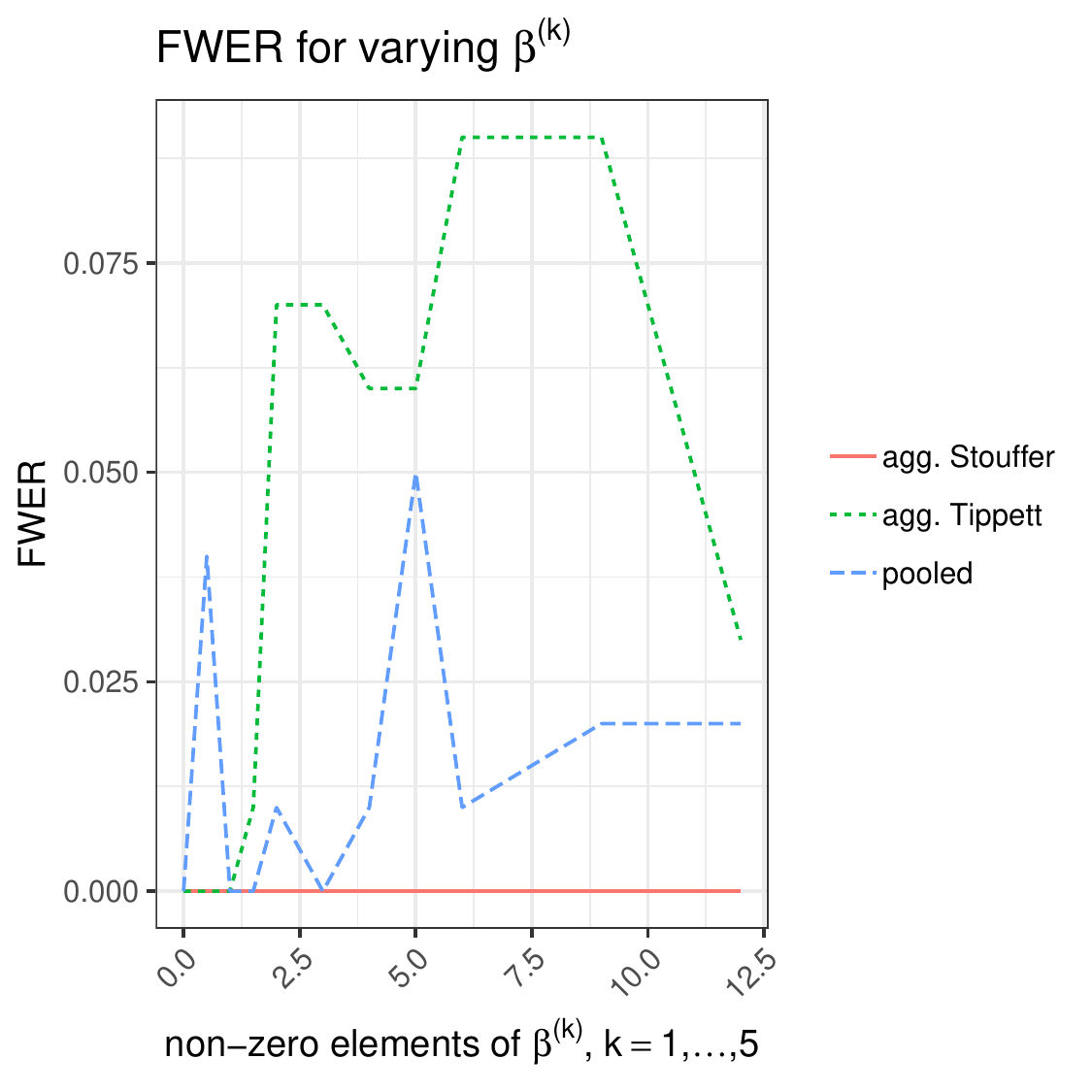}}%
    \caption{Ten datasets: comparison of the adaptive power and FWER of aggregating p-values 
    using Tippett's or Stouffer's rule or by simply 
    pooling multiple studies. 
    The single value of the non-zero elements of $\beta^{(\ell)}$, $\ell
    = 1, \ldots, 5$ is varied, while the non-zero elements of $\beta^{(\ell)}$, 
    $\ell = 6, \ldots, 10$ all take the value 0.5.  
    The common active set is of size 10 and is randomly selected for each
    simulation 
    run. The results are based on 100 simulation runs.}%
    \label{fig8}%
\end{figure}

\begin{figure}%
    \centering
    \subfigure[Adaptive Power.]{\includegraphics[width = 0.45\textwidth]{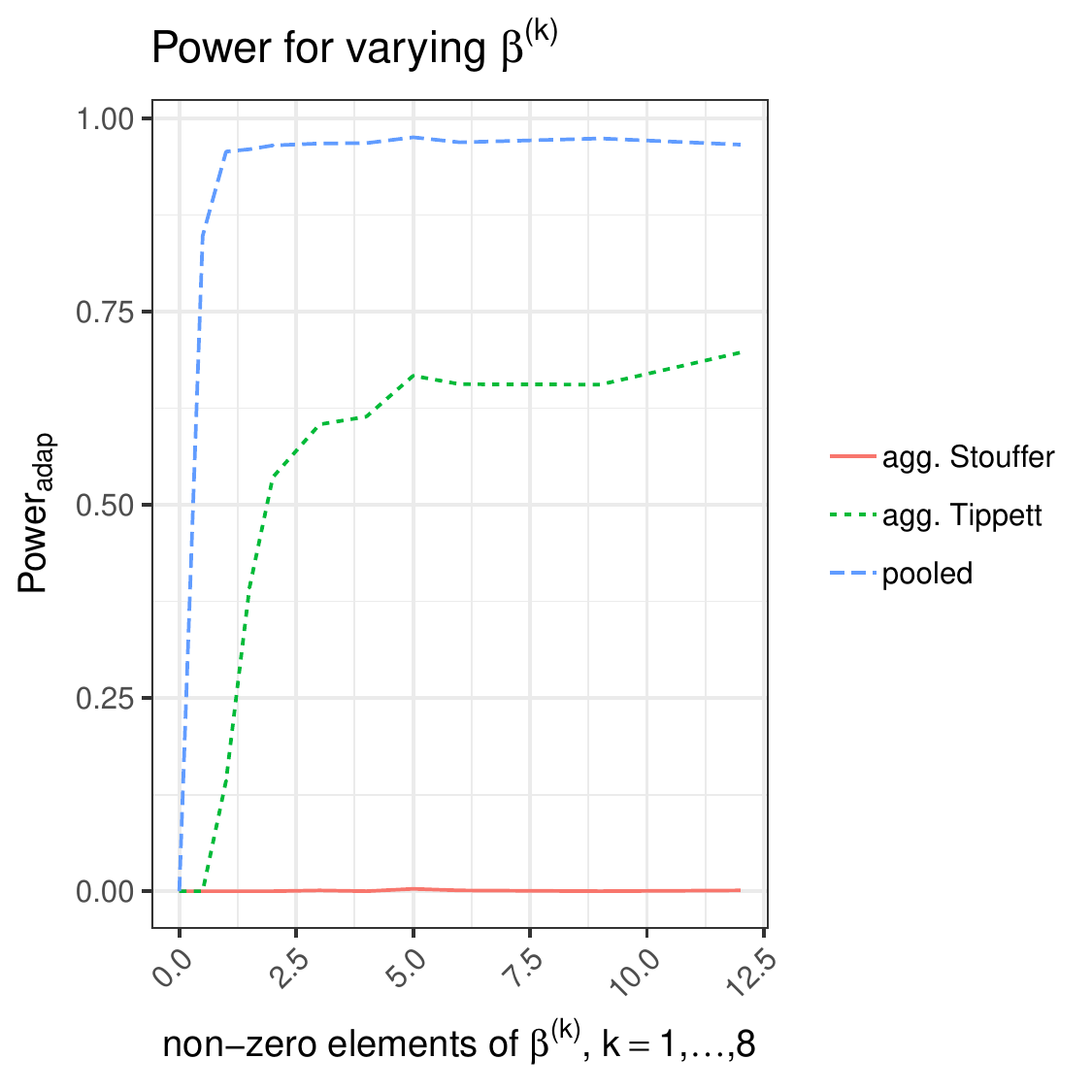}}%
    \qquad
	\subfigure[Family-Wise Error Rate.]{\includegraphics[width = 0.45\textwidth]{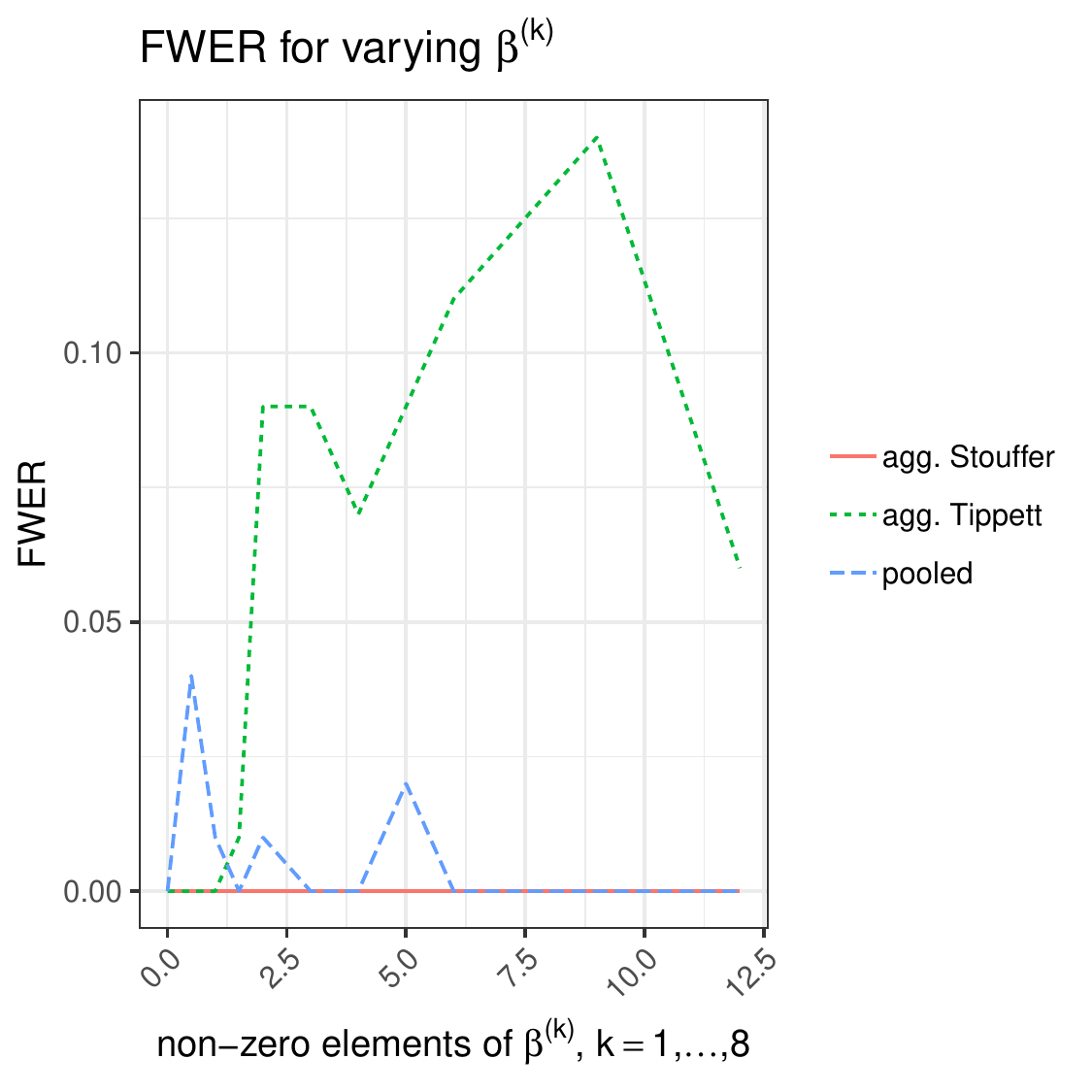}}%
        \caption{Ten datasets: comparison of the adaptive power and FWER of aggregating p-values 
    using Tippett's or Stouffer's rule or by simply 
    pooling multiple studies. 
    The single value of the non-zero elements of $\beta^{(\ell)}$, $\ell
    = 1, \ldots, 8$ is varied, while the non-zero elements of $\beta^{(\ell)}$, 
    $\ell = 9, 10$ both take the value 0.5.  
    The common active set is of size 10 and is randomly selected for each
    simulation 
    run. The results are based on 100 simulation runs.}
    \label{fig9}%
\end{figure}

Aggregating the p-values using Tippett's rule performs 
worse than pooling while aggregation with Stouffer's rule performs
poorly. The results are illustrated in Figures \ref{fig8} and \ref{fig9}.  
The number of observations per dataset is halved compared to the simulation 
in Section \ref{subsec.empiricalmeta} and the number of SNPs is 10 times
larger, both being favourable for pooling. 
We also note that the active sets of the 10 datasets are identical and thus, the
different  
datasets are perhaps still rather ``homogeneous''.
It can be dangerous to pool the datasets because in general there is no
theoretical  guarantee that the FWER is controlled.

\paragraph{Performance of Stouffer's rule.} The main reason why
Stouffer's rule for aggregation of p-values performs so poorly seems to
be the \emph{instability} of the hierarchical inference scheme. For two
datasets having even the same generating distribution, it can easily
happen that
the hierarchical inference scheme provides once a highly significant and
once a non-significant result. And analogously, a similar pattern arises
with more than two datasets. In such situations, Stouffer's rule
performs poorly, as indicated also by Figure \ref{fig.agg}. In the
worst case, if one of the p-values from the different datasets is 1,
then Stouffer's rule won't reject for sure.  

The explanation of the observed instability is as follows. 
The p-values arising from multiple sample splits are aggregated using
\eqref{aggreg} where the correction factor $1/\gamma$ is the price to pay for 
using multi sample splitting. An aggregated p-value can be large or even 1
if a mix of moderate to large (and perhaps also some very few small)
p-values is aggregated.  
Furthermore, the raw p-values of an active cluster can be large or even equal to
1 if the signal is weak or if the selected variables from
Lasso pre-screening have an empty intersection with the cluster of
interest, respectively. The latter issue arises because of the difficulty of
variable screening in very high-dimensional settings with high correlations
among the variables.

\subsection{Theoretical considerations for aggregating p-values and pooling of multiple datasets}

We have illustrated in Figures \ref{fig8} and \ref{fig9} that pooling can 
be clearly better than using Tippett's multiple testing correction.  

To shed some light on the issue, we consider the situation with linear
models as mentioned at the beginning of Section \ref{sec.meta},
\begin{eqnarray*}
  Y_i^{(\ell)} = \sum_{j=1}^p \beta_j^{(\ell)} X_i^{(\ell)} + \eps_i^{(\ell)},\
  i=1,\ldots n_{\ell},
\end{eqnarray*}
over various datasets $\ell = 1\ldots ,m$. For simplicity, we assume that
$n_{\ell} \equiv n$ for all $\ell$ and that the $X_i^{(\ell)}$ are fixed
variables which have been i.i.d. sampled from a distribution with covariance
matrix $\Sigma_X^{(\ell)}$, where $\Sigma_X^{(\ell)} \equiv \Sigma_X$ for all
$\ell$. The latter might be far from being true but our aim here is only to
present a simple argument. 

Consider a statistic for testing $\beta_j^{(\ell)}$: 
\begin{eqnarray*}
  T_j^{(\ell)},\ \,  T_j^{(\ell)} \sim {\cal N}(0,1)\ \mbox{under}\
  \tilde{H}_{0,j},
\end{eqnarray*}
with $\tilde{H}_{0,j}$ as in \eqref{nullhypj}. 
The $t$-test statistic in a linear model satisfies this asymptotically
under mild distributional assumptions on the error term, and
under the assumptions from Section \ref{subsec.multsamplspl}, this
also holds in a sample splitting context as used in our approach and
software. 

Tippett's multiple testing correction \eqref{aggreg-Tippett} is slightly more powerful than
Bonferroni correction, and the latter amounts to consider the maximum of
the test-statistics
\begin{eqnarray*}
  \max_{\ell = 1,\ldots,m} |T_j^{(\ell)}|.
\end{eqnarray*}
It is well known, that due to the Gaussian assumption and under
the null-hypothesis $\tilde{H}_{0,j}$:
\begin{eqnarray*}
  \PP\big[\max_{\ell =1,\ldots ,m} |T_j^{(\ell)}| > \sqrt{c^2 + 2 \log(m)} \ \big] \, \le \,
  2 \exp(-c^2/2).
\end{eqnarray*}
This implies that for $T_j^{(\ell)}$ being the $t$-test statistics for
$\beta_j^{(\ell)}$, the test has power converging to 1 (for any fixed
significance level) if
\begin{eqnarray}\label{detection-tippett}
  \max_{\ell=1,\ldots ,m} \frac{|\beta_j^{(\ell)}|}{\sigma^{(\ell)}
  (\Sigma_X)^{-1}_{jj}} \gg \sqrt{\log(m)/n},
\end{eqnarray}
where $\sigma^{(\ell)} = \sqrt{\Var(\eps^{(\ell)})}$ is the standard deviation
of the noise term $\eps^{(\ell)}$.  
Thus, we see from \eqref{detection-tippett} that Tippett's correction pays a
price with a factor $\sqrt{\log(m)}$, due to multiple testing, instead of
the usual detection rate $1/\sqrt{n}$.

With pooling as described at the beginning of Section 
\ref{subsec.empiricalmeta}, we consider the pooled parameter
in the linear model over all the $m$ datasets:

\begin{eqnarray*}
  \beta^{\mathrm{pool}} = \argmin_{\beta} \EE\Big[ n_{\mathrm{tot}}^{-1}
  \sum_{i=1}^{n_{\mathrm{tot}}} (Y_i - X_i^T \beta)^2 \Big],
\end{eqnarray*}
with corresponding noise term $\eps^{\mathrm{pool}}_i = Y_i - X_i^T
\beta^{\mathrm{pool}}$ and $n_{\mathrm{tot}} = \sum_{\ell=1}^m n_{\ell} = m
n$.  
We then obtain that
\begin{eqnarray*}
  \beta^{\mathrm{pool}} = \sum_{\ell=1}^m \beta^{(\ell)} \PP[Z=\ell],
\end{eqnarray*}
where $Z$ denotes the random variable encoding the index of the
dataset (assuming here a mixture model for the $m$ datasets). In
comparison to \eqref{detection-tippett}, the $t$-test with 
pooled data then leads to the detection
\begin{eqnarray}\label{detection-pooled}
\frac{|\beta_j^{\mathrm{pool}}|}{\sigma^{\mathrm{pool}} (\Sigma_X)^{-1}_{jj}}
  \gg \sqrt{1/(m n)}.
\end{eqnarray}

For comparing \eqref{detection-tippett} with \eqref{detection-pooled}, we
consider two special cases.

\medskip
\emph{Case I (equal $\beta_j$'s).} Suppose that $\beta_j^{(\ell)} \equiv
\beta_j$ for all $\ell$, implying also that the supports of $\beta^{(\ell)}$
are the same. Then it holds that $\beta_j^{\mathrm{pool}} = \beta_j$ and the
detection boundary in \eqref{detection-pooled} is clearly in favor of the pooled
method. This case is ``fairly close'' to the scenario in Figures \ref{fig8} and 
\ref{fig9} where
all the $\beta_j$ just take two values over the $m = 10$ datasets.

\medskip
\emph{Case II (fully distinct supports of $\beta$'s).} Suppose that all the supports of
$\beta^{(\ell)}$ are disjoint and thus: if $\beta_j^{(\ell)} \neq 0$ it must be
that $\beta_j^{\ell'} = 0$ for all $\ell' \neq \ell$. In the balanced case
where $\PP[Z=\ell] \equiv 1/m$, the pooled parameter then equals
$\beta_j^{\mathrm{pool}} = \beta_j^{(\ell)}/m$. Then, the detection boundary
for coefficient $j$ in \eqref{detection-pooled} becomes
\begin{eqnarray}\label{detection-pooled2}
\frac{|\beta_j^{(\ell)}|}{\sigma^{\mathrm{pool}}
  (\Sigma_X)^{-1}_{jj}} \gg \sqrt{m/n},
\end{eqnarray}
and in this case, the Tippett scheme is better (assuming that
$\sigma^{\mathrm{pool}}$ is comparable to $\sigma^{(\ell)}$): compare
\eqref{detection-pooled2} to \eqref{detection-tippett}.

\medskip
The conclusion from this little calculation is as expected, that pooling
can be better than aggregation of p-values if the different datasets
substantially share the supports and signs of the regression coefficients,
as illustrated in Figures \ref{fig8} and \ref{fig9}. In general, including 
e.g. different
covariances of the covariates, pooling can be inadequate and is exposed to
a misspecified model. Thus, Tippett's aggregation of p-values is the safer
procedure (and e.g. Stouffer's aggregation rule is not really a
  competitor in our setting with multi sample splitting for hierarchical
  testing, as pointed out in the paragraph at the end of Section
  \ref{subsec.empiricalmeta.m10}).

\subsection{Software for aggregating p-values of multiple studies} \label{subsec.multiplestudies}

It is very convenient to combine the information of multiple studies by 
aggregating p-values as described in Section \ref{sec.meta}.  
The package \texttt{hierinf} offers two methods for jointly estimating 
a single hierarchical tree for all datasets using either of the
functions  
\texttt{cluster\char`_var} or \texttt{cluster\char`_position}; compare with 
Section \ref{subsec.clustering}. Testing is performed by the function 
\texttt{test\char`_hierarchy} in a top-down manner given by the joint 
hierarchical tree. For a given cluster, p-values are calculated based on the 
intersection of the cluster and each dataset (corresponding to a study) and 
those p-values are then aggregated to obtain one p-value per cluster using 
either Tippett's rule \eqref{aggreg-Tippett} or Stouffer's method
\eqref{aggreg-Stouffer}; see argument \texttt{agg.method} of the function 
\texttt{test\char`_hierarchy}. The difference and issues of the two methods for 
estimating a joint hierarchical tree are described in the following two 
paragraphs.

\medskip 

The function \texttt{cluster\char`_var} estimates a hierarchical tree based on 
clustering the SNPs from all the studies. Problems arise if the
studies do not measure the same SNPs and thus,
some of the entries of the dissimilarity matrix cannot be calculated. 
By default, pairwise complete observations for each pair of SNPs are
taken to construct the dissimilarity matrix. 
This issue affects the building of the hierarchical tree but the testing of
a given cluster remains as described before.  

The function \texttt{cluster\char`_position} estimates a hierarchical tree based
on the genomic positions of the SNPs from all the studies. 
The problems mentioned above do not show up here since SNPs, maybe
different ones for various datasets, can still be uniquely assigned to
genomic regions.  

\medskip

The only difference in all the function calls is that the arguments \texttt{x}, 
\texttt{y}, and \texttt{clvar} are now each a list of matrices instead of 
just a single matrix.  Note that the order of the list elements of the arguments
\texttt{x}, \texttt{y}, and \texttt{clvar} matter, i.e. the user has to
stick to the order that the first element of the three lists corresponds to
the first dataset, the second element to the second datasets, and so on.  
One would replace the corresponding element of the list containing the
control covariates (argument \texttt{clvar}) by \texttt{NULL} if some dataset 
has no control covariates.
If none of the datasets have control covariates, then one can simply omit the 
argument. Note that the argument \texttt{block} defines the second level of the 
tree which is assumed to be the same for all datasets or studies.
The argument \texttt{block} has to be a \texttt{data.frame} which contains 
all the column names (of all the  datasets or studies) and their assignment 
to the blocks. The aggregation method can be chosen using the 
argument \texttt{agg.method} of the function \texttt{test\char`_hierarchy}, 
i.e. it can be set to either \texttt{"Tippett"} or \texttt{"Stouffer"}. 
The default aggregation method is Tippett's rule \eqref{aggreg-Tippett}.

The example below demonstrates the functions \texttt{cluster\char`_var} and
\texttt{test\char`_hierarchy} for two datasets / studies measuring the 
same SNPs.

\begin{lstlisting}[language=R]
# The datasets need to be stored in different elements of a list. 
# Note that the order has to be the same for all the lists. 
\end{lstlisting}

%\newpage

\begin{lstlisting}[language=R]
# As a simple example, we artificially split the observations of the 
# toy dataset in two parts, i.e. two datasets. 
set.seed(89)
ind1 <- sample(1:500, 250)
ind2 <- setdiff(1:500, ind1)
\end{lstlisting}

%\newpage

\begin{lstlisting}[language=R]
sim.geno.2dat  <- list(sim.geno[ind1, ], 
                       sim.geno[ind2, ])
sim.clvar.2dat <- list(sim.clvar[ind1, ], 
                       sim.clvar[ind2, ])
sim.pheno.2dat <- list(sim.pheno[ind1], 
                       sim.pheno[ind2])    
\end{lstlisting}

% \newpage

\begin{lstlisting}[language=R]
# Cluster the SNPs
dendr <- cluster_var(x = sim.geno.2dat, 
					 block = block, 
					 # the following arguments have to be specified 
					 # for parallel computation
					 parallel = "multicore",
					 ncpus = 2)
					 
# Test the hierarchy using multi sample split
set.seed(1234)
result <- test_hierarchy(x = sim.geno.2dat, 
						 y = sim.pheno.2dat, 
						 clvar = sim.clvar.2dat,
						 dendr = dendr, 
						 family = "binomial",
						 # the following arguments have to be 
						 # specified for parallel computation
						 parallel = "multicore",
						 ncpus = 2)
\end{lstlisting}
The above \textsf{R} code is evaluated in parallel; compare with Section 
\ref{subsec.parallel} for more details about the software for parallel 
computing.

%\newpage

The output shows one significant group of SNPs and one single SNP.
\begin{lstlisting}[language=R]
> print(result, n.terms = 4)
  block   p.value    significant.cluster                         
1 chrom 1 NA         NA                                          
2 chrom 2 0.02659100 SNP.532, SNP.721, SNP.882, SNP.520, ... [15]
3 chrom 2 0.01100256 SNP.993     
\end{lstlisting}
The significance of a cluster is based on the information of both datasets. 
For a given cluster, the p-values of each dataset were aggregated using 
Tippett's rule as in \eqref{aggreg-Tippett}. Those aggregated p-values are 
displayed in the output above. We cannot judge which dataset (or both or 
combined) inherits a strong signal such that a cluster is shown significant 
but that is not the goal. The goal is to combine the information of multiple 
studies.

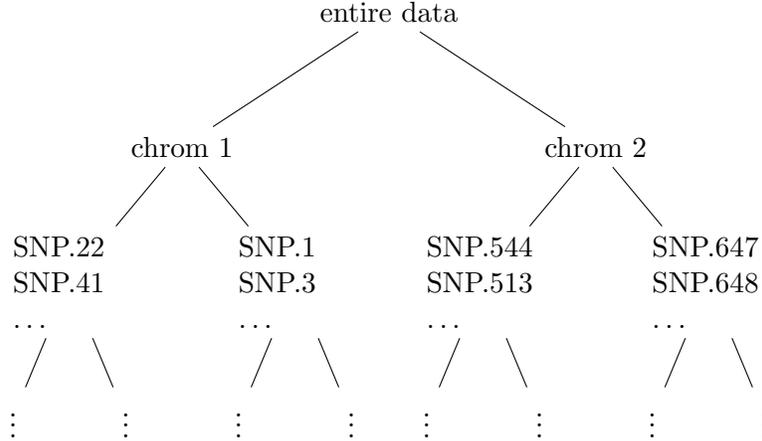
\begin{figure}[!htb]
\begin{center}
\begin{tikzpicture}[level distance=1.8cm,
  level 1/.style={sibling distance=5.5cm},
  level 2/.style={sibling distance=3cm}, 
  level 3/.style={sibling distance=1.5cm}] 
\node (z) {entire data} 
  child {node (a) {chrom 1}
    child {node (aa) [text width=1.5cm]{SNP.22 SNP.41 $\ldots$}
        child {node (aaa) {$\vdots$}}
        child {node (aab) {$\vdots$}}}    	
    child {node (ab) [text width=1.5cm]{SNP.1 SNP.3 $\ldots$}
        child {node (aaa) {$\vdots$}}
        child {node (aab) {$\vdots$}}}
  }
  child {node (b) {chrom 2}
    child {node (ba) [text width=1.5cm]{SNP.544 SNP.513 $\ldots$}
            child {node (aaa) {$\vdots$}}
            child {node (aab) {$\vdots$}}}
    child {node (bb) [text width=1.5cm]{SNP.647 SNP.648 $\ldots$}
            child {node (aaa) {$\vdots$}}
            child {node (aab) {$\vdots$}}}
  };
\end{tikzpicture}
\caption{Illustration of a possible single jointly estimated tree for 
multiple studies based on clustering the SNPs. The second level of the 
hierarchical tree is defined by chromosome 1 and 2 (defined by the argument 
\texttt{block} of the functions \texttt{cluster\char`_var} / 
\texttt{cluster\char`_position}). The function \texttt{cluster\char`_var} / 
\texttt{cluster\char`_position} builds a separate 
hierarchical tree for each of the chromosomes.}\label{fig7} 
\end{center}
\end{figure}

The crucial point is that the testing procedure goes top-down through a single 
jointly estimated tree for all the studies and only continues if at least one 
child is significant (based on the aggregated p-values of the multiple datasets) 
of a given cluster. The algorithm determines where to stop and naturally 
we get one output for all the studies. A possible single jointly estimated tree 
of the above \textsf{R} code is illustrated in Figure \ref{fig7}. In our 
example, both datasets measure the same SNPs. If that would not be the case, 
then intersection of the cluster and each dataset is taken before calculating 
a p-value per dataset / study and then aggregating those.

\section{Discussion and conclusions}

We provide a review of hierarchical inference for high-dimensional
(generalized) linear models, particularly aiming for the analysis of
genome-wide association studies (GWAS) where the dimensionality is in the order
$O(10^6)$ and sample size typically in the thousands. Inferring statistical
significance in such high-dimensional settings is very challenging: we
believe that \emph{hierarchical} inference is a very natural and powerful
approach towards better and more reliable inference in GWAS. Obviously, multiple
datasets or studies contain more information. We advocate the use of
meta-analysis within a single hierarchical structure which is simple and
coherent. 

Our new implementation in the \textsf{R}-package \texttt{hierinf} provides
many possibilities: two options for constructing hierarchical structures,
fitting linear and logistic linear response models with possible additional
adjustment for external control variables, and efficient parallel
computation. Our software is a major cornerstone for enabling the practical
use of hierarchical inference for GWAS, controlling the FWER. A different
way of performing hierarchical inference can be done within the framework
of selective inference for controlling the FDR or a conditional FDR
\citep{brzyski17,helleretal2017}. 

Many open problems remain. Among them, we name here a few. (1) The issue of
hidden confounders: even when taking all measured SNPs into the analysis,
unobserved confounding can lead to spurious and wrong associations. An
extreme example is given by \citet{Novembreetal08}, and mixed models
\citep{raketal13,zhouetal13} may only 
account in part for hidden confounders. (2) Another point is the
debate whether the familywise error rate (FWER) is a too strict criterion
to work with, in contrast to the false discovery rate (FDR): the FWER is
simpler to control, especially in hierarchical and closed testing
schemes. We refer also to \citet{goeman2011} for an interesting discussion
on this point. In the classical non-hierarchical inference, the ranking about
significance of single hypotheses is not influenced whether the user
chooses adjustment of p-values with the Bonferroni-Holm or the
Benjamini-Hochberg procedure to control the FWER or FDR, respectively. In the
hierarchical case though, this remains unclear. In addition, the p-values
from multi-sample splitting, as used in our procedure and software, might
be unreliable: it is challenging, in particular for logistic regression
\citep{sur18}, to come up with reliable and powerful p-values for testing
single or groups of regression 
coefficients which are reliable and powerful in high-dimensional
settings. (3) The role of the hierarchy is an issue of power, as long as we
assume fixed design and a correct model specification. The FWER
control holds for any fixed hierarchical structure, but the power
typically depends on the chosen hierarchy. 
We have not
considered here the region-based approach from \citet{meijer15} which allows
for a supervised form of the groups or clusters at the price of a more severe
multiple testing correction: in absence of a broad comparison, we do not want to
give general recommendations. Our view is (in part) to enable the users to
try our approach and make their own judgment: our \textsf{R}-software
package \texttt{hierinf} should provide substantial support to do so. 

\section*{Acknowledgments}
We thank the Ricardo Cao for arranging the possibility present our results
as a discussion paper. We also thank two anonymous reviewers for very insightful
and helpful comments.
 
\bibliographystyle{apalike} 
\bibliography{references}

\end{document}